\documentclass[10pt]{iopart}
\pdfoutput=1
\usepackage{subfigure}
\usepackage{amssymb}
\usepackage{graphicx}
\usepackage{bm}
\usepackage{color}

\graphicspath{ {./Figures/} }

\usepackage[backend=biber,citestyle=numeric,sorting=none]{biblatex}
\begin{document}

\title{Simultaneous iterative learning control of mode entrainment and error field}
\author{W.~Choi and F. A.~Volpe}

\address{Dept of Applied Physics and Applied Mathematics, Columbia University, New York, NY 10027, USA}

\begin{abstract}
It is shown that static error fields (EFs) can severely limit the maximum rotation frequency achievable in mode entrainment by applied rotating fields. 
It is also shown that the rotation non-uniformities caused by an EF can be used to diagnose and correct said EF in real time. 
Simulations using typical DIII-D data show that this can be achieved within a small number of mode rotation periods by an iterative learning control algorithm. 
In addition to rapidly correcting the EF, this gives access to high entrainment frequencies that would 
not be accessible otherwise, and paves the way to rotational stabilization and improved mode control. 
\end{abstract}

\ioptwocol

\section{Introduction}
Error field (EF) correction and mode entrainment are inter-related topics of 
great importance for the stable, high-confinement operation of tokamaks.

The importance and benefits of EF correction are well-known and extensively reviewed  \cite{Reimerdes2011, Strait2014}.
When left uncorrected, intrinsic EFs can lead to a braking of plasma rotation, the formation of magnetic islands, a loss of confinement and possibly a disruption.

\begin{figure}[t]
       \includegraphics[scale=0.6,trim={0 0 0.2cm 0},clip]{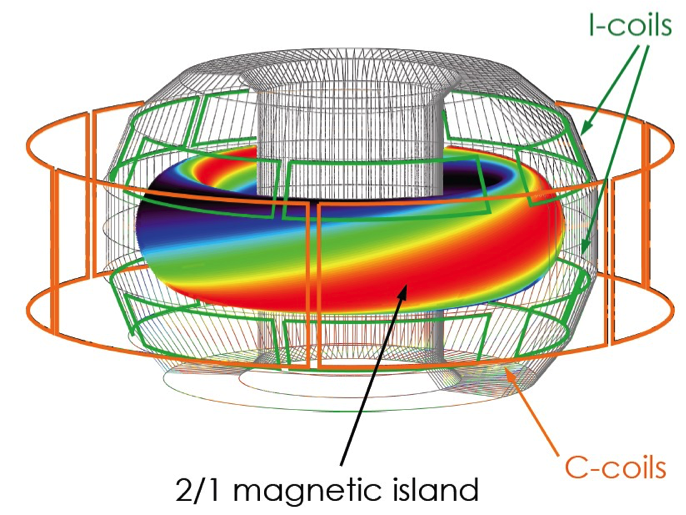}
       \caption{Perturbed current pattern, sinusoidal in helical angle \(m\theta - n\phi\), of a \(m/n=2/1\) magnetic island mapped onto a thin surface.
		Also depicted are the internal I-coils (green) and external C-coils (red) used to generate 3D fields on DIII-D.} 
		\label{fig:islandExample}
\end{figure}

The toroidal rotation of magnetic islands can be sustained by means of ``mode entrainment'' with applied rotating fields \cite{Morris1990}. 
Such fields, like the static perturbations for EF correction, 
are exerted by non-axisymmetric control coils internal or external to the vessel, similar to 
DIII-D's I- and C-coils depicted in figure \ref{fig:islandExample}.
Entrainment is important because it prevents mode locking \cite{Volpe2009} 
and can assist the island stabilization by modulated Electron Cyclotron Current Drive \cite{Choi2018}. 
In addition, mode entrainment at up to 300 Hz has led to the recovery of a pedestal in the edge pressure profile at DIII-D, for reasons yet to be understood \cite{Shiraki2013}. 
Related to that, and as a possible interpretation of those observations, rapid entrainment is 
expected to stabilize the mode, either by flow shear effects \cite{Chen1990, Lahaye2009} or due to the interaction with the resistive wall \cite{Brennan2014}.
Entrainment is also important for diagnostic purposes, as it allows to characterize a mode by 
toroidally steering it in front of toroidally localized diagnostics.  

EFs are theoretically \cite{Fitzpatrick1993} and experimentally \cite{Volpe2009} 
known to affect mode entrainment. 
In particular, the mode rotates non-uniformly if the applied rotating 
Resonant Magnetic Perturbation (RMP) used for entrainment is stronger than the uncorrected EF 
(\(|B_{RMP}| > |B_{EF}|\))\cite{Volpe2009}. 
Here resonant means having the same toroidal mode number \(n\). 
Rotations are incomplete if \(|B_{RMP}| \le |B_{EF}|\) \cite{Volpe2009}. 
More generally, EFs affect the rotation of magnetohydrodynamic (MHD) modes in general (not restricted 
to islands), whether sustained by applied rotating fields or by other torques. 
This implies that the motion of MHD modes can be used as a tool to diagnose the EF. 
This has been accomplished for kink modes and tearing modes on EXTRAP-T2R \cite{Volpe2013,Sweeney2016}, and 
for locked or nearly locked tearing modes on DIII-D \cite{Shiraki2014,Shiraki2015}. 

In those previous works, though, the EF was diagnosed {\em after} the plasma discharge, and corrected in the following shot. 
In the present paper, instead, 
we show that it is possible to iteratively use the mode rotation (including an initially incomplete rotation) as a tool to diagnose the EF {\em in real time}.  
Each iteration corrects the EF and thus make the mode rotation more uniform, 
which results in an even more precise EF 
estimate and EF correction, even more uniform rotation, and so forth. 
By this method, EFs are characterized and corrected in as little as 4 mode rotation periods. 
This is faster than previous works and much faster than traditional ``compass scans'', 
requiring 4 plasma discharges \cite{Reimerdes2011}. 
Once the EF has been made sufficiently small and the mode entrainment sufficiently uniform, 
one can increase the rotation frequencies to values that were not accessible before. 
In other words, mode entrainment is used to rapidly characterize the EF, and improved EF correction is used for more uniform, faster entrainment.
Faster entrainment paves the way to rotational stabilization, improved mode control and reliable disruption avoidance. 

The remainder of this paper is structured as follows: section \ref{sec:IdentifyEF} discusses how the error field is measured.
The Iterative Learning Control (ILC) concept is introduced in section \ref{sec:ILC} and ILC simulations 
are shown in section \ref{sec:simResults}.
Section \ref{sec:conclusions} summarizes the major conclusions and implications of this work.

\section{Identifying error field}
\label{sec:IdentifyEF}
An accurate measurement of the EF is required in order to achieve good correction.
This section details how the EF's perturbative effect on an entrained island's rotational motion can be used to diagnose the amplitude and phase of the EF.

\subsection{Static or slowly rotating island}
In a basic model, a locked mode (LM) is expected to align with the vector-sum of the EF and applied RMP. 
Namely, the toroidal phase is expected to equal the phase of the vector-sum of the RMP and EF field:
\begin{equation}
\phi_{LM} = \angle \vec{B}_{RMP}+\vec{B}_{EF}.
\label{eqn:vectorSum}
\end{equation}  
This holds true when neglecting the effect of eddy currents in the wall and consequent ``wall torque'' 
acting on the island (which is a legitimate approximation for low frequencies), as 
well as other torques, imparted for example by the rotating 
plasma which the island is partly frozen in. Here ``low frequencies'' refers to lower than the inverse wall time.

\begin{figure}[t]
      \includegraphics[width=0.5\columnwidth]{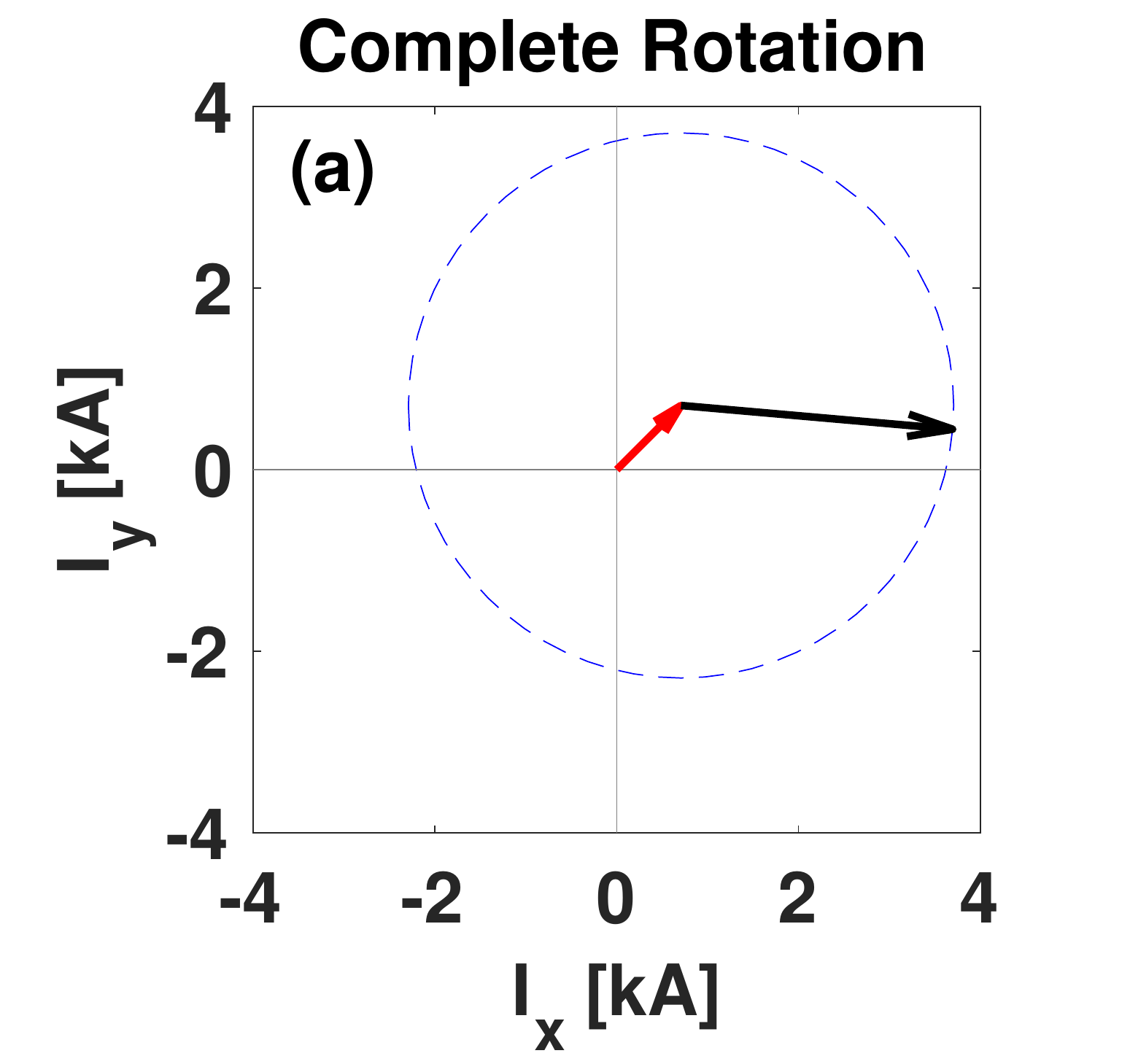}        
      \includegraphics[width=0.5\columnwidth]{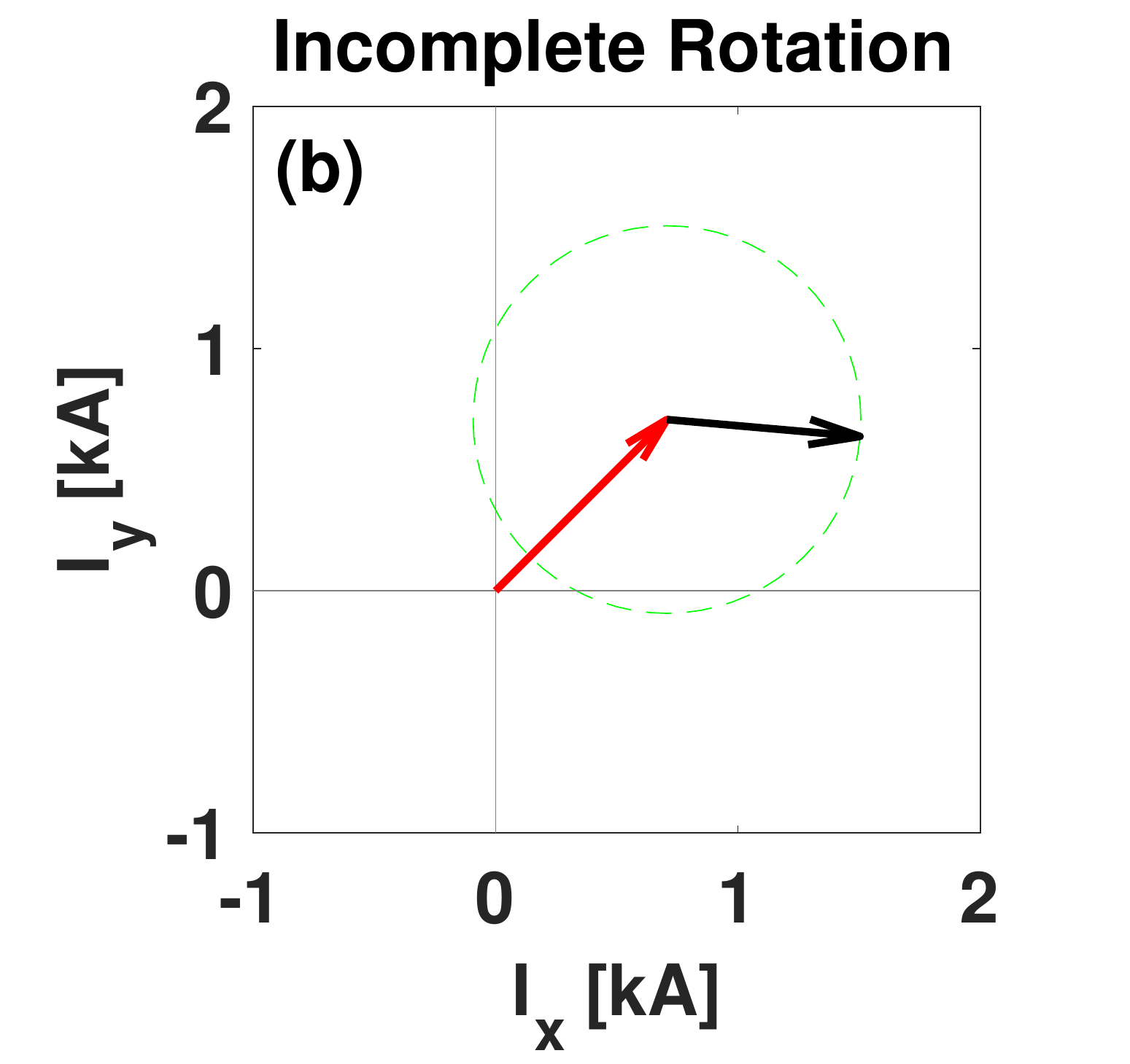}
      \includegraphics[width=\columnwidth]{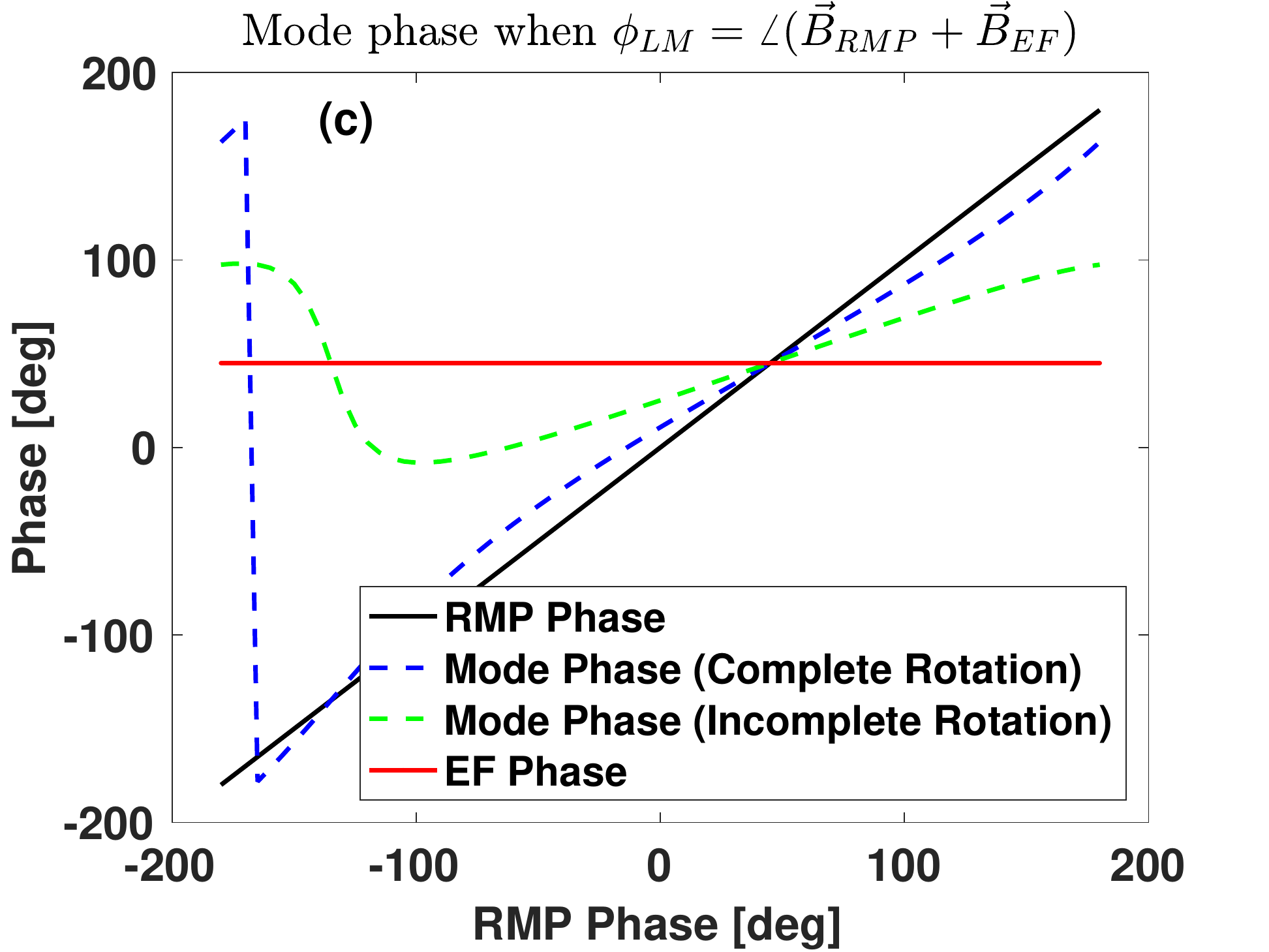}
		\caption{Resultant locked mode phase for a slowly rotating applied RMP that is (a) larger or (b) smaller than the EF. 
		Here, \(I_{x, y}\) are the x- and y-components of applied RMP current and equivalent EF current.  
		(c) shows the deviation in mode phase compared to the applied RMP phase if the RMP is larger (blue) or smaller (green) than the EF.}
		\label{fig:VecPlots}
\end{figure}

Figure \ref{fig:VecPlots} illustrates the vector-sum of a fixed EF (red arrow) and rotating RMP (black arrow).
Depending on whether the applied RMP magnitude is greater or smaller than the EF magnitude,
the resultant vector (thus, the LM) performs full rotations (figure \ref{fig:VecPlots}a), or oscillates 
between two values of \(\phi_{LM}\) (figure b).
Note that, in the first case, as the RMP rotates uniformly, 
the resultant rotates non-uniformly.

Both cases are accurately described by
\begin{equation}
\phi_{LM}(t) = \arctan \left[\frac{B_{EF,y} +B_{RMP,y}(t)}{B_{EF,x} +B_{RMP,x}(t)}\right]
\label{eqn:arctan}
\end{equation}
where the subscripts $_x$ and $_y$ denote the $x$ and $y$ components in a frame whose $z$ axis is aligned with the axis of 
symmetry of the tokamak. 
Figure \ref{fig:VecPlots}c presents plots of $\phi_{LM}(t)$ 
for the two cases. Again, in one case $\phi_{LM}$ spans 
all values, in the other it spans a limited range. 

Note that $B_{RMP,x}(t)$ and $B_{RMP,y}(t)$ are known and 
$\phi_{LM}(t)$ can be measured. 
Hence, equation \ref{eqn:arctan} appears to offer a direct calculation of the two unknown EF components based on just two 
measurements of $\phi_{LM}$ at different times. 
With noise included, however, this fitting function performs poorly when only two data-points are used, and more data is needed\textemdash from at least half a rotation period (Sec.\ref{subsec:ItPeriod}). Fitting over a longer time-interval improves the quality of the fit, but reduces the time-resolution 
of the EF identification. 

\begin{figure}[t]
	\includegraphics[width = 0.9 \columnwidth]{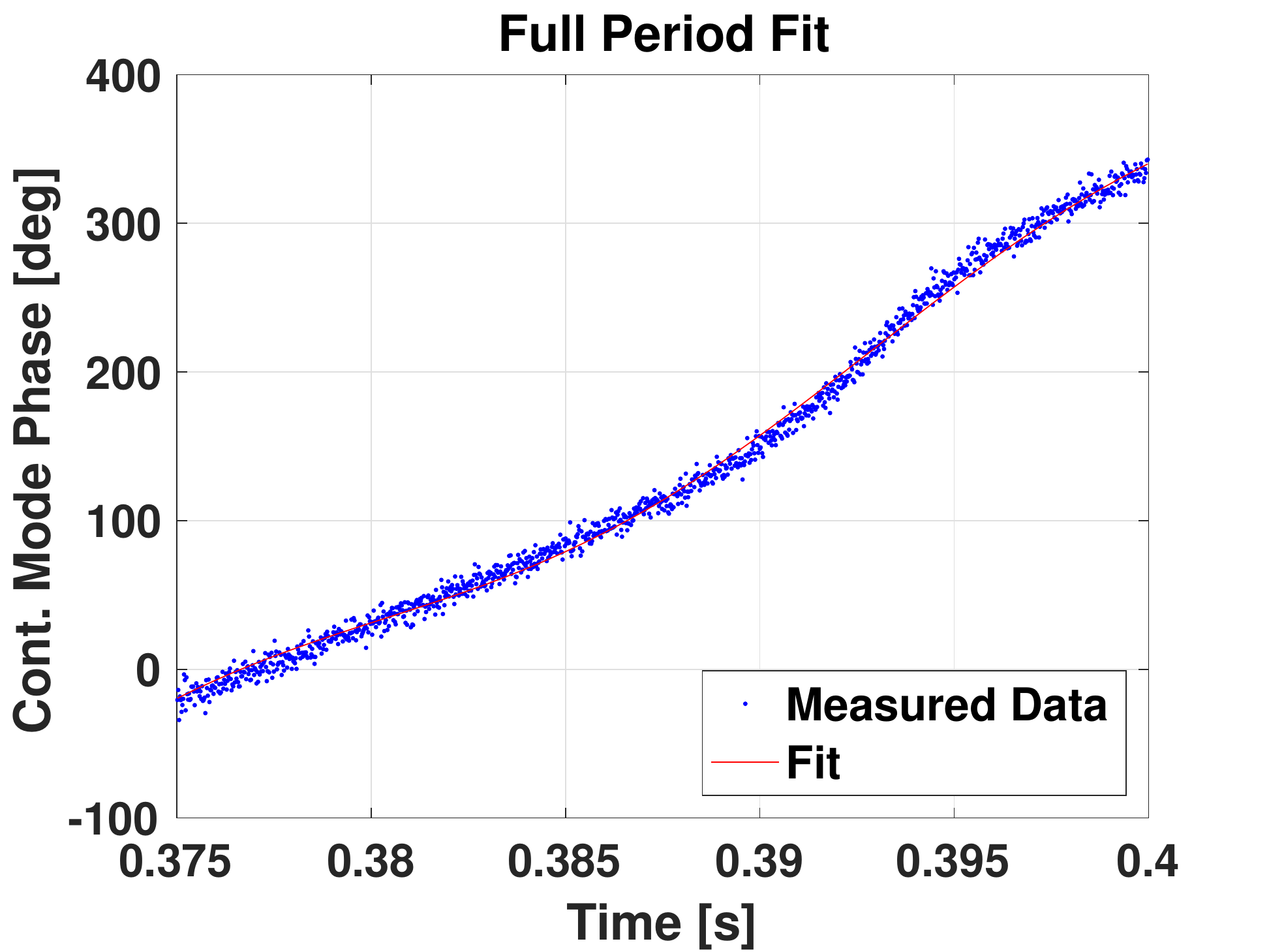}
	\caption{An example of the linear + sinusoidal fitting function performing well on noisy mode phase data.}	
	\label{fig:FullPeriodFit}
\end{figure}

Normally the phase $\phi_{LM}$ in equation \ref{eqn:arctan} 
is defined as taking values between 
-180\(^\circ\) to 180\(^\circ\). 
However, phase ``jumps'' from -180\(^\circ\) to 180\(^\circ\), or vice versa, are problematic to the fitting of equation \ref{eqn:arctan}. 
Thus, a continuous, unraveled  $\phi_{LM}$ was used for the fitting instead.

If the rotation is complete and just slightly non-uniform 
(due to a small EF), the fitting function can be replaced by an \(n=1\) sinusoidal disturbance on top of a straight, constant frequency entrainment:
\begin{equation}
	\phi_{LM}(t) = mt + b + a sin(\omega t + c)
	\label{eqn:fittingFunction}.
\end{equation}
Here \(m\) and \(\omega\) are fixed by the frequency of rotation, \(b\) is an offset, and \(a\) and \(c\) are used to calculate the EF amplitude and phase respectively. $b$ and $c$ are related 
to the LM and EF phases, respectively, relative to the 
origin of the toroidal coordinate. 

As for incomplete rotation, the amplitude of the EF can be derived from the spread of the locked mode phase 
(figure \ref{fig:VecPlots}),
and the EF phase is simply the mean value of a full period of mode phase.

\subsection{Considerations for fast rotation} \label{subsec:ConsidFast}
It was previously stated that these equations are only accurate at low frequency with respect to the inverse wall time, which on DIII-D is of the order of 50~Hz.
For higher frequencies, eddy currents induced in the wall affect the results in several significant ways.

First and foremost, a rotating island induces currents in the wall that {\em drag} upon the island itself. 
Thus, in order to achieve torque balance, the phase of the applied RMP must necessarily {\em lead} the mode phase.
The difference between the two depends on the desired frequency, wall frequency, mode amplitude and RMP amplitude.

Additionally, due to wall shielding, the RMP effectively applied at the mode location is reduced in amplitude. 
It also lags in phase when compared to the RMP applied by the coils.

The situation is further complicated with the inclusion of the error fields.
When an uncorrected EF is added to an applied RMP of  constant amplitude and frequency, 
the island will accelerate or decelerate as it approaches 
alignment with the EF or departs from it. 
The resultant motion of the island will necessarily be oscillatory (super-imposed to island rotation or not, 
depending on $|B_{RMP}|$ exceeding $|B_{EF}|$ or not). 
The oscillatory motion of the island will induce time-dependent currents in the wall, causing further non-uniformities in the island motion. 
 
Given the above considerations, equation \ref{eqn:vectorSum} is 
only valid in the low-frequency limit (much lower than 50 Hz, at 
DIII-D). 
If that estimate is used at higher frequencies and without corrections for wall effects,  there will likely be a small but non-zero residual EF that can  continue to prevent smooth entrainment.

\section{Iterative learning control}
\label{sec:ILC}
Iterative learning control (ILC) is a type of tracking control best suited for repetitive situations \cite{Moore1999, Felici2015, Ravensbergen2018}.
For locked mode entrainment, both the requested mode phase and the phase perturbation caused by the EF are periodic in nature, 
making this an appropriate application for ILC.

\begin{figure}
	\includegraphics[width = 0.9 \columnwidth]{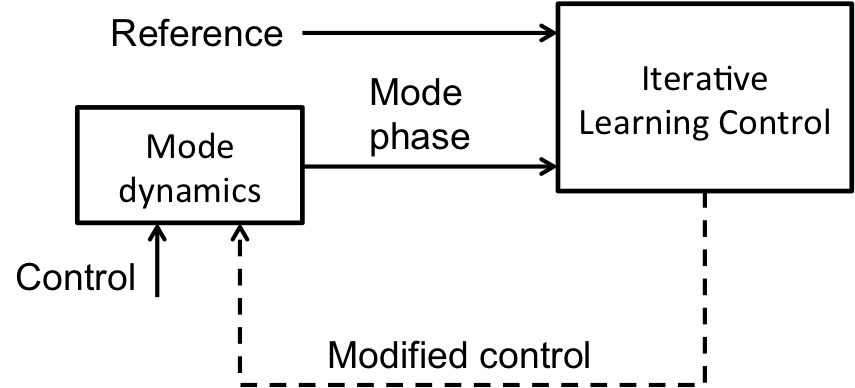}
	\caption{A schematic of the iterative learning controller. After each iteration, the control input is modified to improve mode phase trajectory.}
	\label{fig:ILCSchematic}
\end{figure}

Figure \ref{fig:ILCSchematic} depicts the basic principles of an ILC designed for simultaneous 
mode entrainment and error field correction (EFC).
Starting with a simple applied RMP rotating at constant frequency and amplitude, the mode phase will have some non-uniformity due to the error field.
The ILC uses this information to quantify the EF, and apply the optimal correction to zero the measured EF.
This modified control is added to the initial command, and the next iteration will have improved smoothness of rotation.
The ILC retains the corrective behavior ``learned'' in previous iterations, and improves over time.

One advantage of ILC is the capability to accept rough estimates of EF instead of requiring precise measurements.
As discussed in the previous section, wall shielding and other effects reduce the accuracy of EF measurements using equation \ref{eqn:vectorSum}.
However, the ILC is able to use just the first order estimate of the EF within each iteration, 
improving its correction over several cycles and eventually approaching the exact currents needed to zero the EF.

\section{Simulation results}
\label{sec:simResults}
This section demonstrates the capabilities of ILC using the cylindrical model described in \cite{Olofsson2016}.
This time-dependent code treats the island as a rigid body with a helical current pattern associated to it, rotating in the presence of a conductive wall. 
Realistic representation of DIII-D's I-coils and typical error field values are included. 

\subsection{Entrainment limits without controller} 
A sufficiently strong rotating RMP can entrain a locked mode despite induced currents in the wall resisting the motion and 
despite the EF effects discussed in section 
\ref{subsec:ConsidFast}. 
Equivalently, for a given RMP strength there is a maximum 
frequency at which 
torque balance can be established. This ``critical entrainment frequency''   
depends on the various parameters affecting the RMP, EF and wall 
torques (and other torques considered, if any). 
Only few parameters, however, change significantly from one 
DIII-D discharge to the other and cause major changes in torques. The most prominent is the island width $w$. 

\begin{figure*}[t]
	\includegraphics[width = 0.33\textwidth]{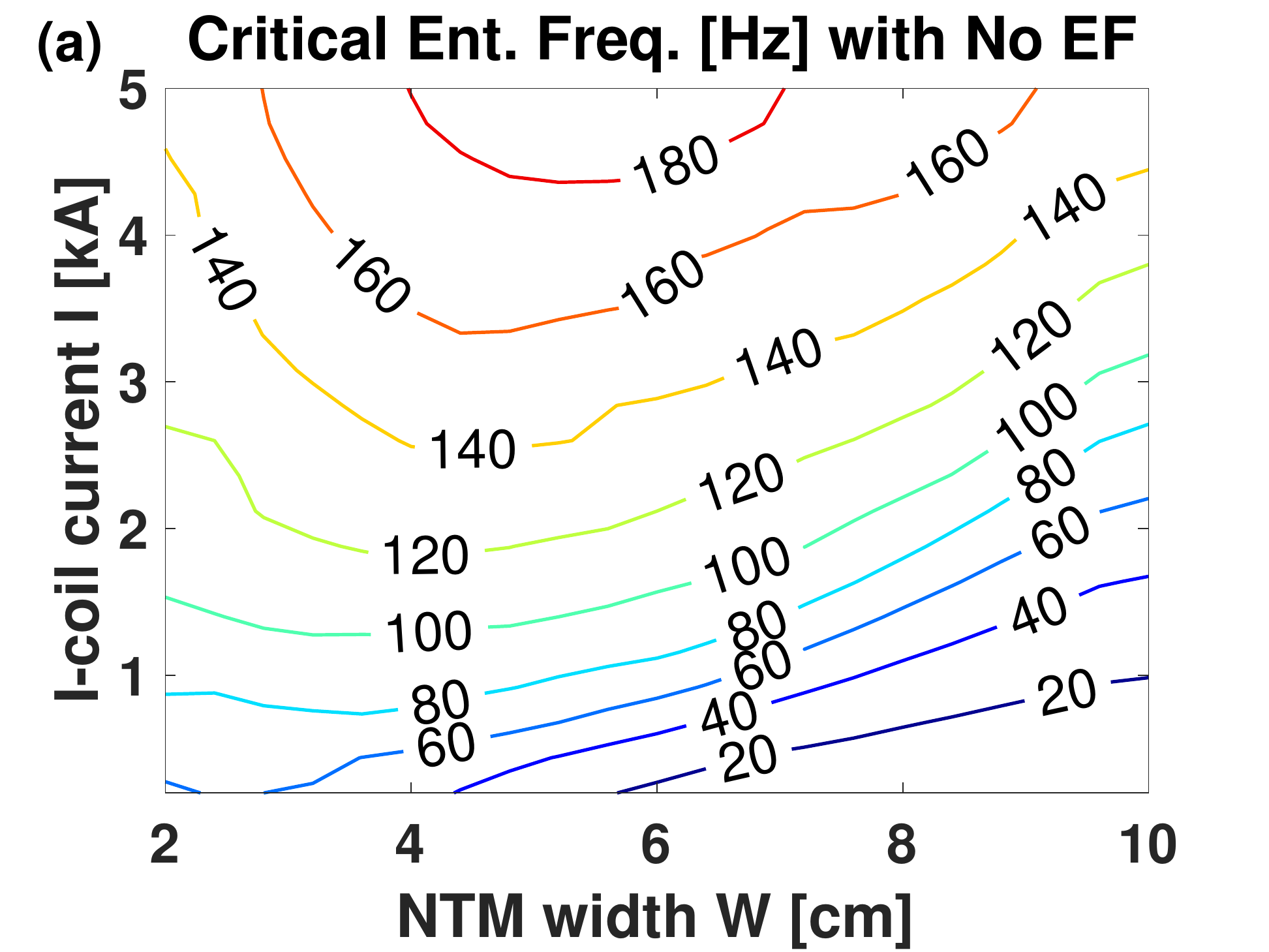}
	\includegraphics[width = 0.33 \textwidth]{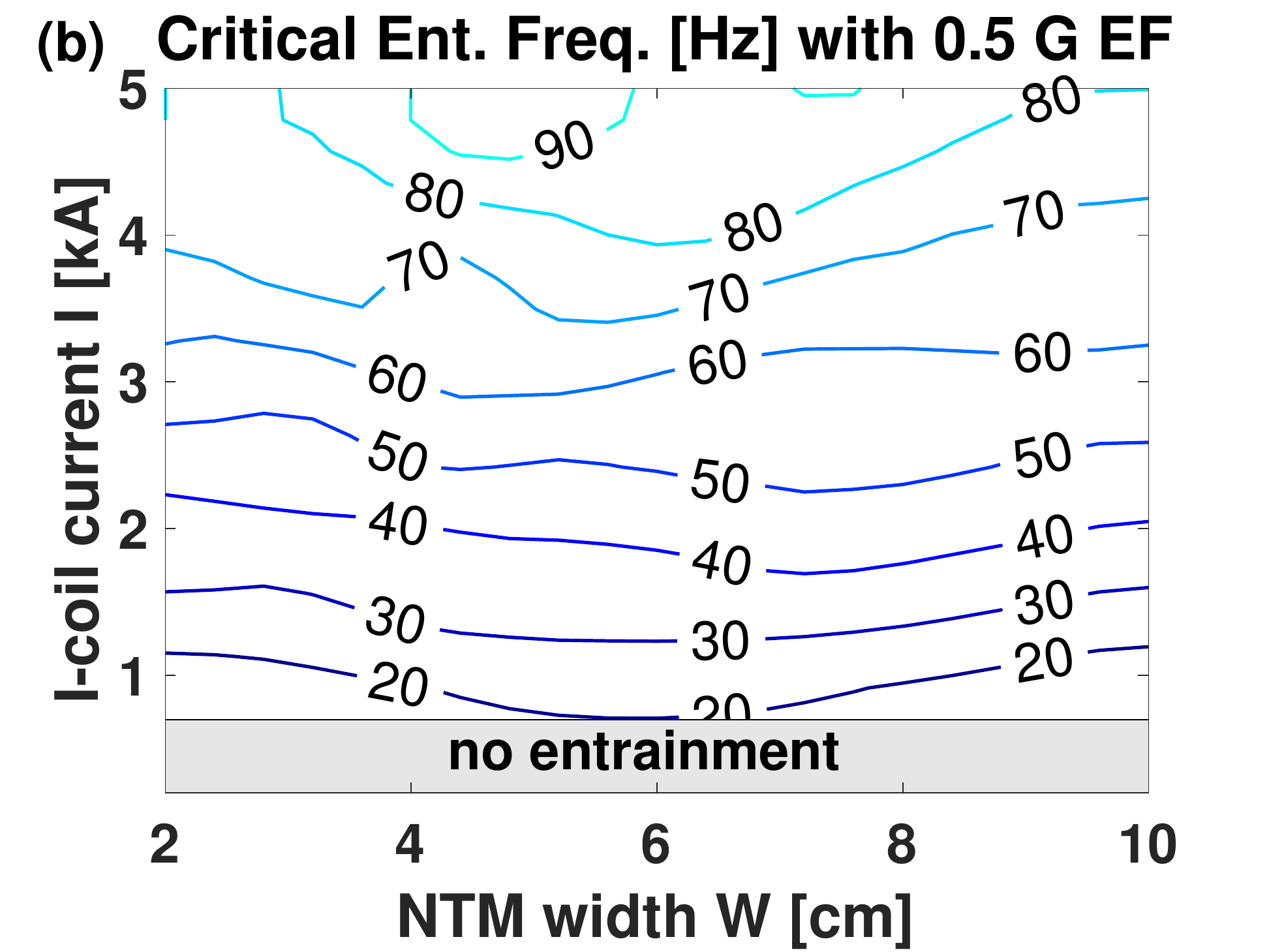}
	\includegraphics[width = 0.33 \textwidth]{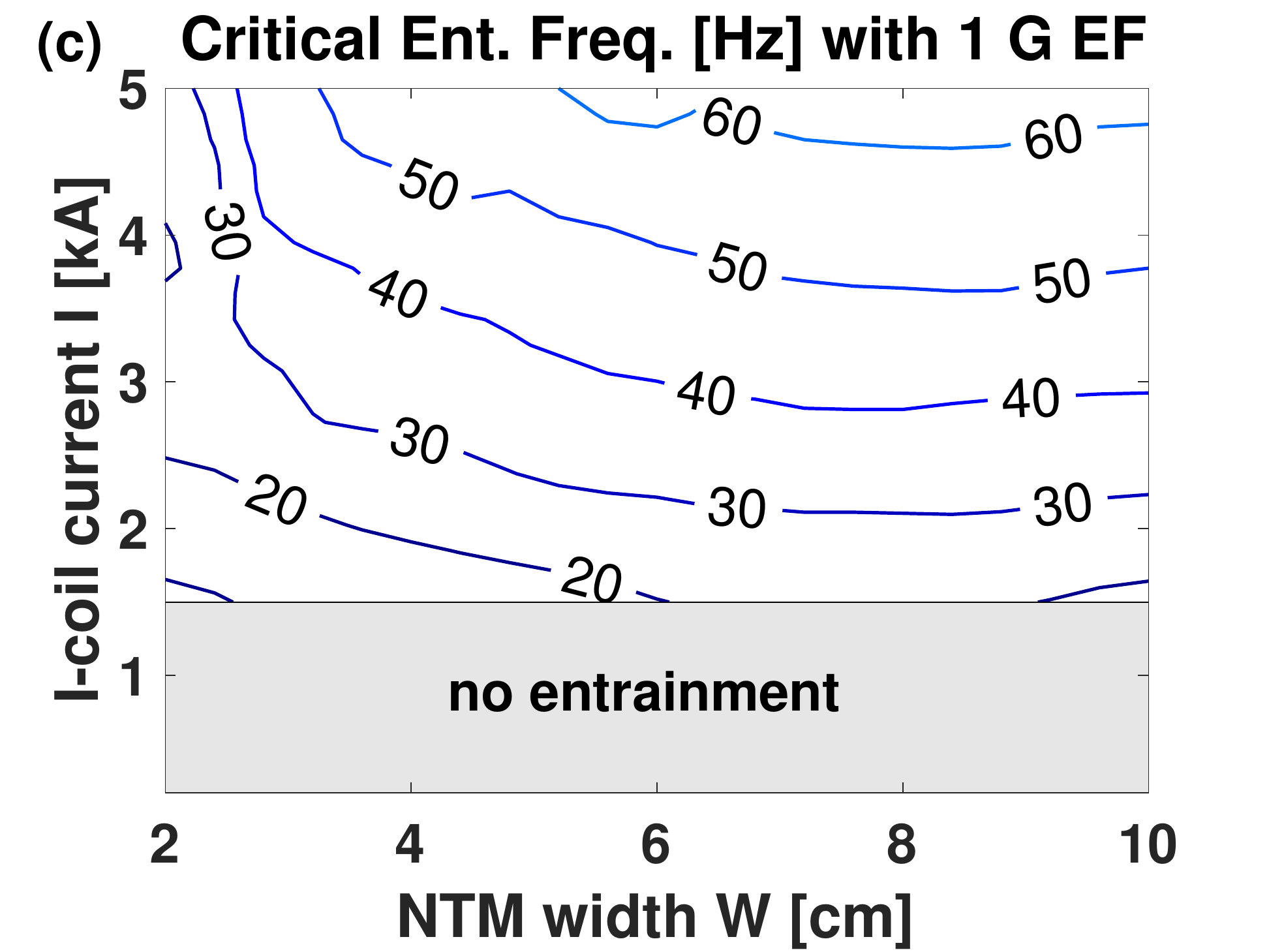}
	\caption{Critical entrainment frequency limits (a) without EF and with EFs of (b) 0.5G and (c) 1G. It is clear that as EF amplitude increases, the maximum frequency for stable entrainment is significantly reduced.}
	\label{fig:critFreqContours}
\end{figure*}

The critical frequency is contour-plotted in figure \ref{fig:critFreqContours} as a function of $w$ and of the RMP strength. 
Figure \ref{fig:critFreqContours}a shows the maximum frequency at which the shielded rotating RMP and wall-drag can achieve torque balance in absence of any EF, previously reported in \cite{Olofsson2016}.
Figure \ref{fig:critFreqContours}b and c, on the other hand, show the maximum frequency at which {\em stable} entrainment 
can be established in the presence of a small (0.5~G) and typical, uncorrected (1~G) DIII-D error field, respectively. 
This frequency is defined as the limit beyond which the combination of EF and induced wall-currents makes the entrainment unstable, in the following sense: 
the instantaneous rotation frequency $f$ varies, 
within a rotation period, 
by a large amount that diverges with $f$.
The entraining frequency at which such divergence occurs is taken to be the critical frequency. 
This and other definitions of entrainment loss or far-from-ideal entrainment are discussed in the Appendix.

Figure \ref{fig:critFreqContours} confirms that, as the EF amplitude increases, the minimum current at which entrainment is possible increases, as expected.
The regions labeled ``no entrainment'' were estimated based on a conversion of using roughly 1.3~kA RMP to correct a 1~G EF.
The parabolic shape of the ``no EF'' contours (figure \ref{fig:critFreqContours}a) is a result of the RMP torque and wall torque varying as $w^2$ and $w^4$, respectively. 
It is worth noting that contours with EF ``flatten the parabola'' and are less sensitive to island width. 
As the EF increases, it is evident that the ratio of \(|B_{RMP}|\) and \(|B_{EF}|\) plays the dominant role in determining the entrainment frequency limit.
Thus, the EF must be corrected well below the opertationally accessible RMP amplitude in order to achieve high frequency entrainment.

\subsection{Time-dependent simulation of ILC}
A typical simulation is set up as follows: 
until \(t=400\)~ms, the fixed-width island is entrained to a 
constant amplitude rotating RMP, in the presence of a prescribed EF. 
The iterative learning controller is activated at 400~ms.
Using the previous rotation period as input, 
the EF is estimated and partly corrected by d.c. offsets in the 
I-coil currents (those coils carry also the 
a.c. currents that exert the rotating RMP). 
Each iteration thereafter has a length of two rotation periods: 
the first to apply the EFC and allow the mode rotation to settle, the second to sense the mode rotation and deduce the EF.

\begin{figure}[t]
	\includegraphics[width = 0.8 \columnwidth]{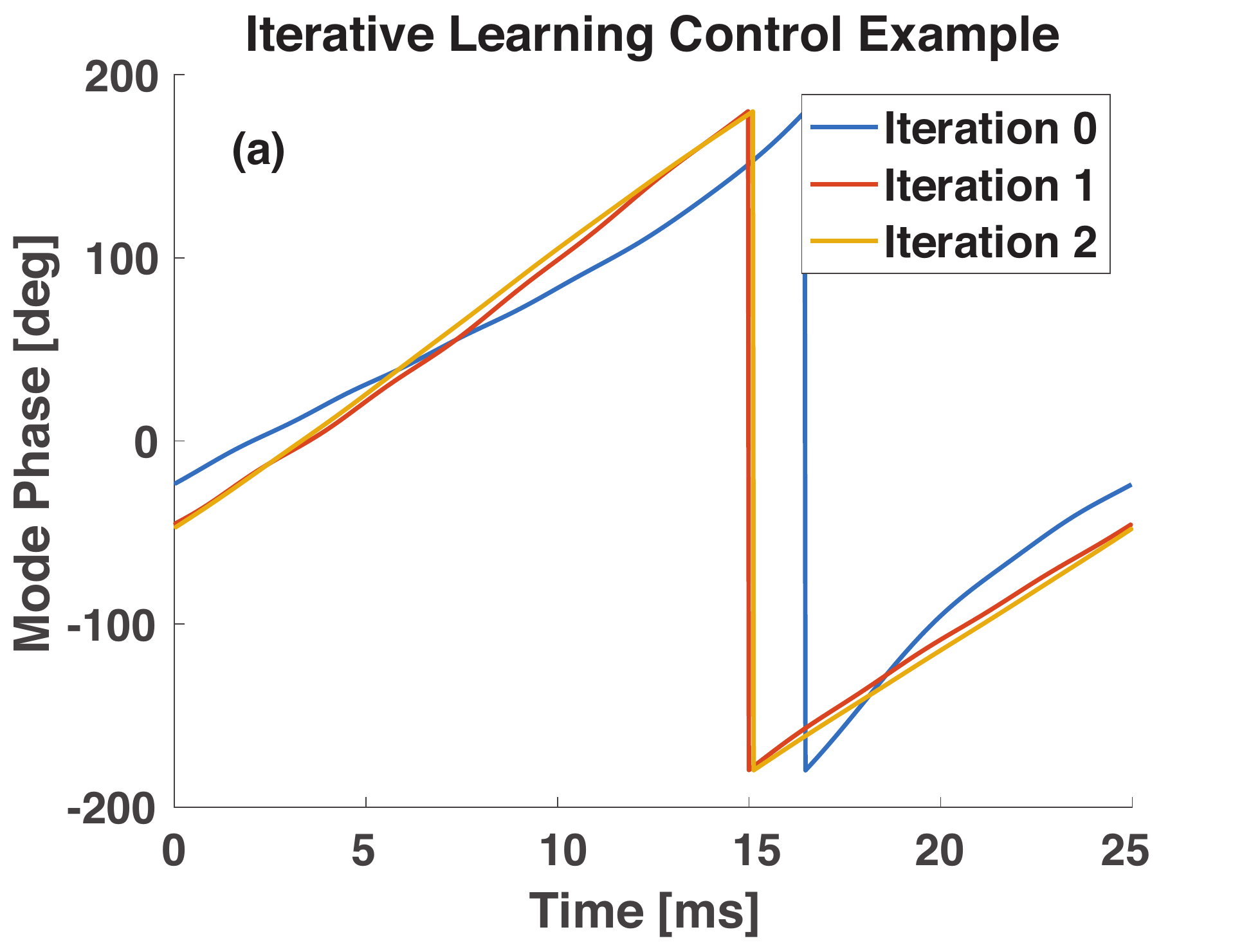}
	\includegraphics[width = 0.8 \columnwidth]{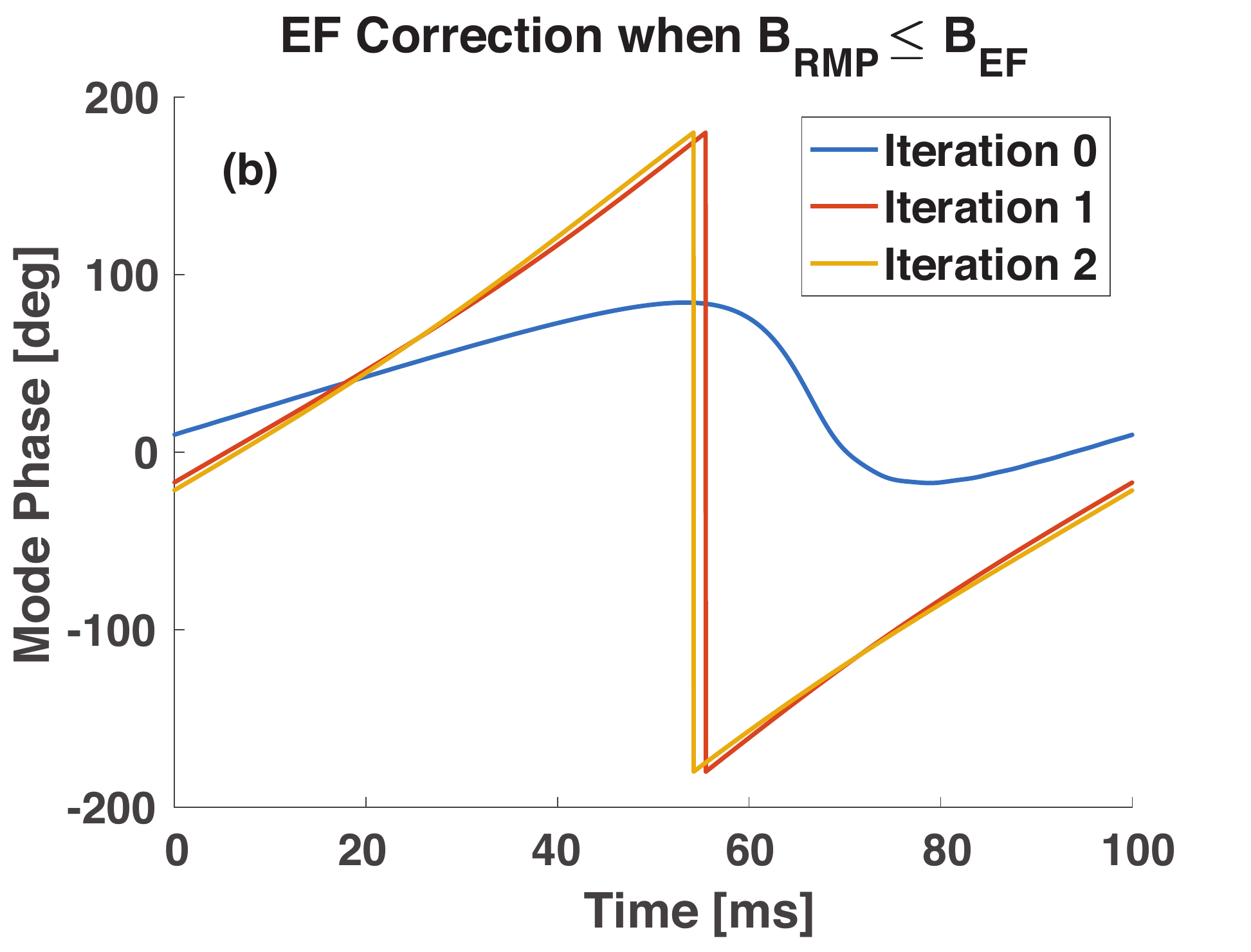}
	\caption{Simulated mode rotation in response to ILC in the presence of (a) 3~kA RMP rotating at 40~Hz correcting a 1~G EF or (b) 0.8~kA, 10~Hz RMP to correct a 1~G EF. Only two iterations are shown for brevity.}
	\label{fig:SingleSimulation}
\end{figure}

Let us consider a 3~kA, 40~Hz RMP and 1~G EF. 
Figure \ref{fig:SingleSimulation}a shows that in this case 
the EF is corrected in about two iterations. 
Iteration 0 refers to the rotation period immediately preceding 400~ms, and measurements from subsequent iterations are shifted backwards for direct comparison. 

Figure \ref{fig:SingleSimulation}b is an example of ILC correctly diagnosing the EF despite the applied RMP being small, 
resulting in incomplete rotation.
After the first iteration, the EF is sufficiently well-corrected for the rotations to become complete, and the ILC converges quickly thereafter.

\begin{figure}[t]
	\includegraphics[width = 0.8 \columnwidth]{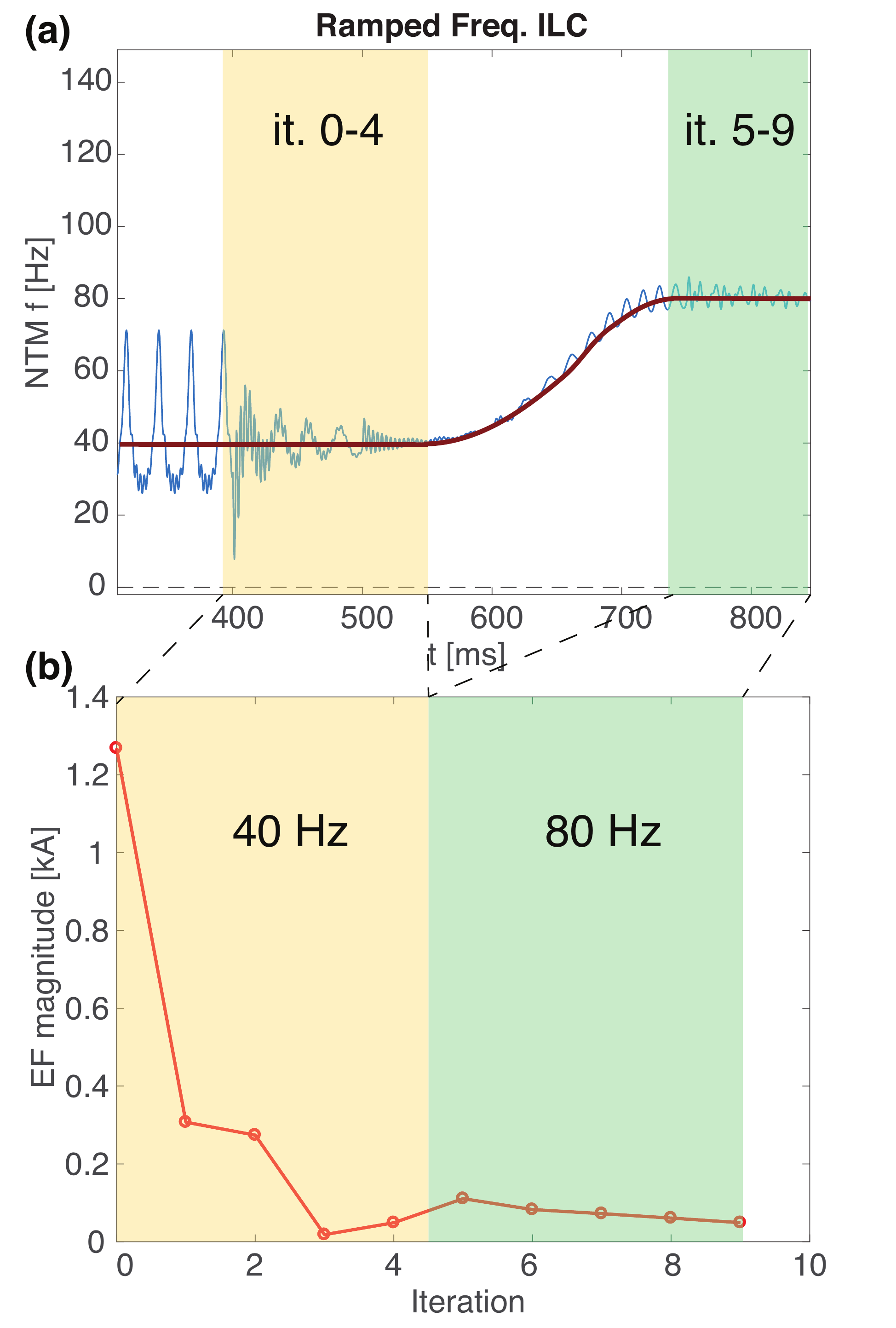}
	\caption{Ramped frequency method: (a) required (red) and actual (blue) mode rotation frequency for an initial entrainment at 40~Hz and simultaneous EF correction followed by a ramp to 80~Hz. (b) Corresponding evolution of 
	the residual EF.}
	\label{fig:RampedFreqILC}
\end{figure}

Entrainment at higher frequency is possible when the EF is sufficiently reduced.
One method to achieve this is through a ramped frequency.
Within the same discharge, one can start operating the 
ILC at lower frequency, to then increase the frequency after a few iterations, after the EF has been reduced. 
The example in figure \ref{fig:RampedFreqILC} consists of 4 iterations at 40~Hz followed by a frequency ramp and another 4 iterations at 80~Hz. 
A direct attempt at 80~Hz entrainment would have resulted in completely inaccurate estimates of EF and failure to entrain, 
as 80 Hz exceeds the maximum entrainment frequency 
for that island width and the initial, uncorrected EF 
(figure \ref{fig:critFreqContours}).  
Subsequent ``corrections'' would have driven the system into dramatic  oscillations in island motion and even more far-fetched 
estimates of EF.

\subsection{Controller robustness}
\begin{figure}[t]
	\includegraphics[width = 0.8 \columnwidth]{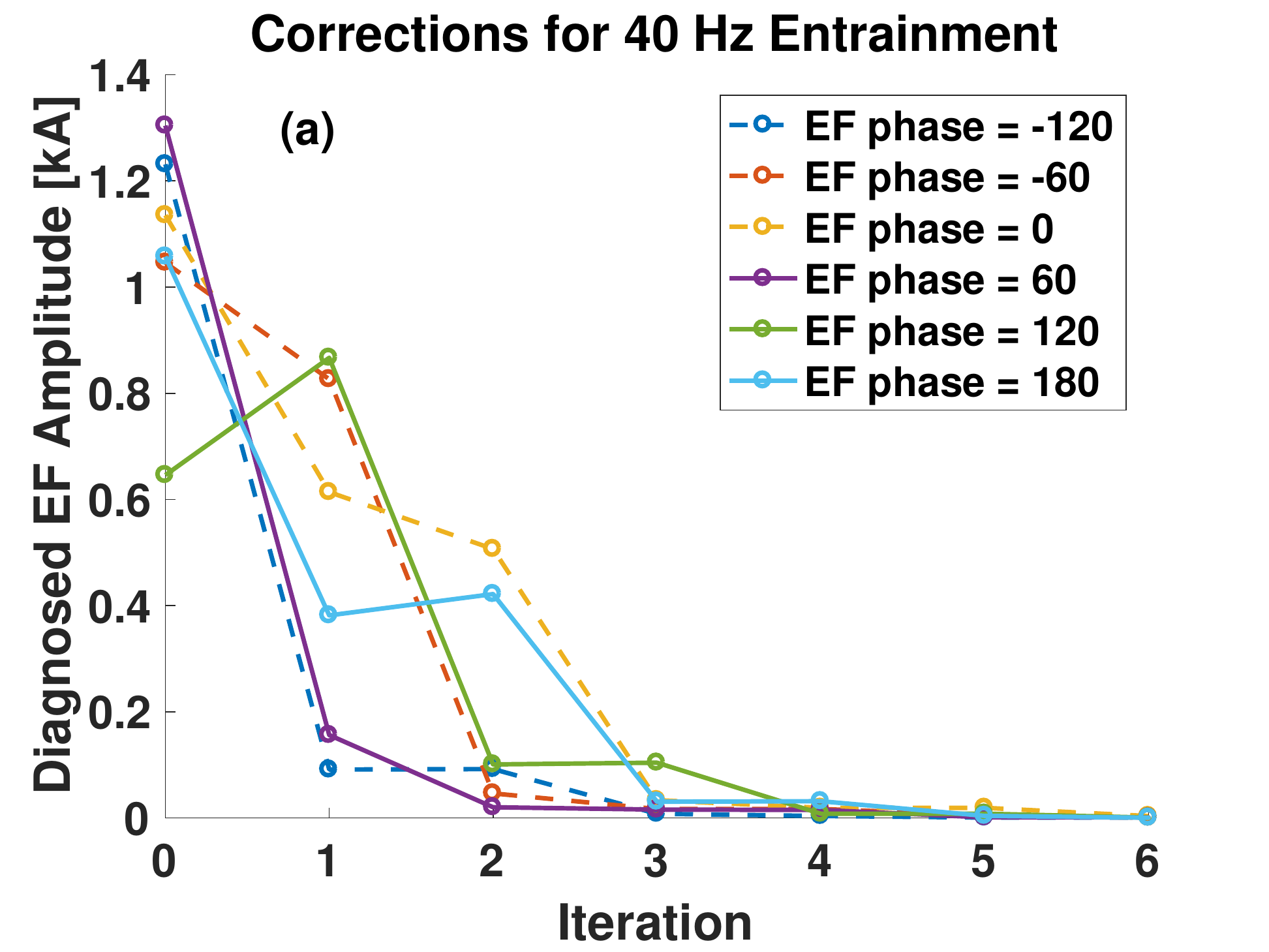}
	\includegraphics[width = 0.8 \columnwidth]{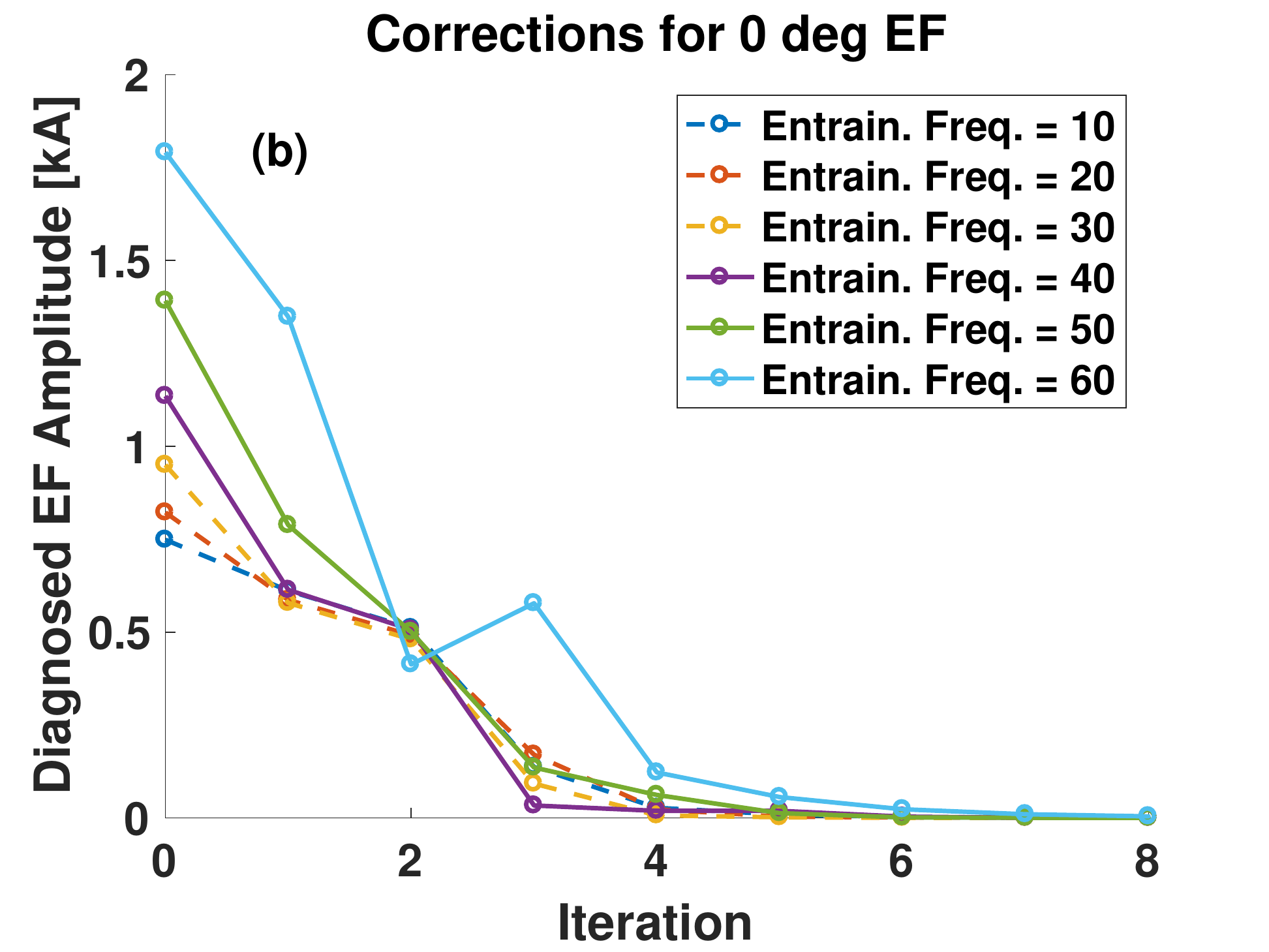}
	\includegraphics[width = 0.8 \columnwidth]{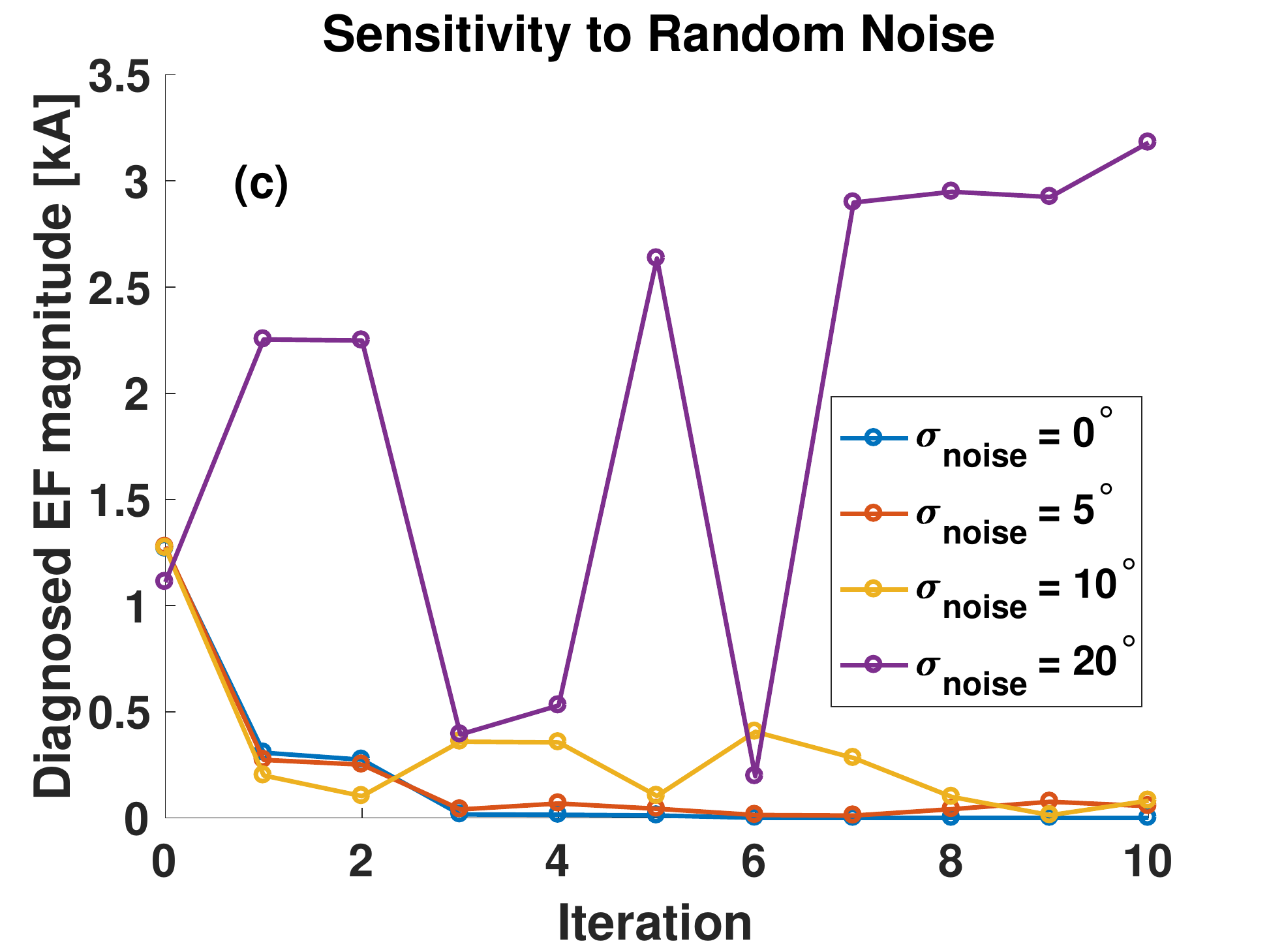}
	\caption{Multiple ILC runs showing robustness and convergence for different (a) initial EF phase, (b) entrainment frequency, and (c) noise.}
	\label{fig:ILCConvergence}
\end{figure}

Robustness to a variety of initial conditions is an essential feature for any controller.
Figure \ref{fig:ILCConvergence} shows that the ILC converges quickly for different EF amplitudes and phases (panel a) and entrainment frequencies (panel b).
It should be noted that a 40~Hz entrainment frequency implies a 50~ms iteration period, which means that most cases achieve convergence in 200~ms (4 iterations).

A well-designed controller should also be robust against random noise in the $\phi_{LM}$ measurement.
When a normally distributed noise\textemdash with standard deviation of \(5^\circ\)\textemdash is added to the measurement, the ILC still converges to the expected value within the first few iterations.
Incidentally, \(5^\circ\) is a realistic phase uncertainty for medium-to-large islands on DIII-D.  
However, if the noise-level exceeds a threshold located between \(10^\circ\) and \(20^\circ\) (orange and purple 
lines in figure~\ref{fig:ILCConvergence}c), 
the ILC converges initially to the correct EF estimate but eventually to an incorrect value, due to poor fitting of noisy phase data. Consequently, the applied EFC is incorrect and the EF is not zeroed. 
In fact, in the \(20^\circ\) 
case in figure~\ref{fig:ILCConvergence}c, the effective EF actually increases, causing an even bigger rotation non-uniformity. 
It is then important to distinguish between these EFs misdiagnosed from noisy data and actual changes in the intrinsic EF (caused for example by changes in plasma conditions or coil-currents).

Note that a sufficiently large EF (that causes significant non-uniformity in rotation) can be correctly inferred 
from the fit, even when the phase data are noisy. The real issue is with small residual EFs: in that case the ILC 
can incorrectly estimate the EF.  
One solution is thus to add the following logic to the controller: an additional rotation-period can be used 
for a second fitting; 
any true changes in EF amplitude or phase should persist between the two measurements, while false EFs fitted from noise would not be repeatable. 


\subsection{Iteration period}
\label{subsec:ItPeriod}
The first version of the ILC uses two rotation periods per iteration.
The reasoning is that the first period allows the motion to settle after applying the correction, and the second period is used as input for the next iteration.
This helps to prevent a feedback scenario where the applied correction currents affects mode rotation, 
this temporary deviation is incorrectly attributed to the EF, and the next iteration will apply an undesired correction.

\begin{figure}
	\includegraphics[width = 0.8 \columnwidth]{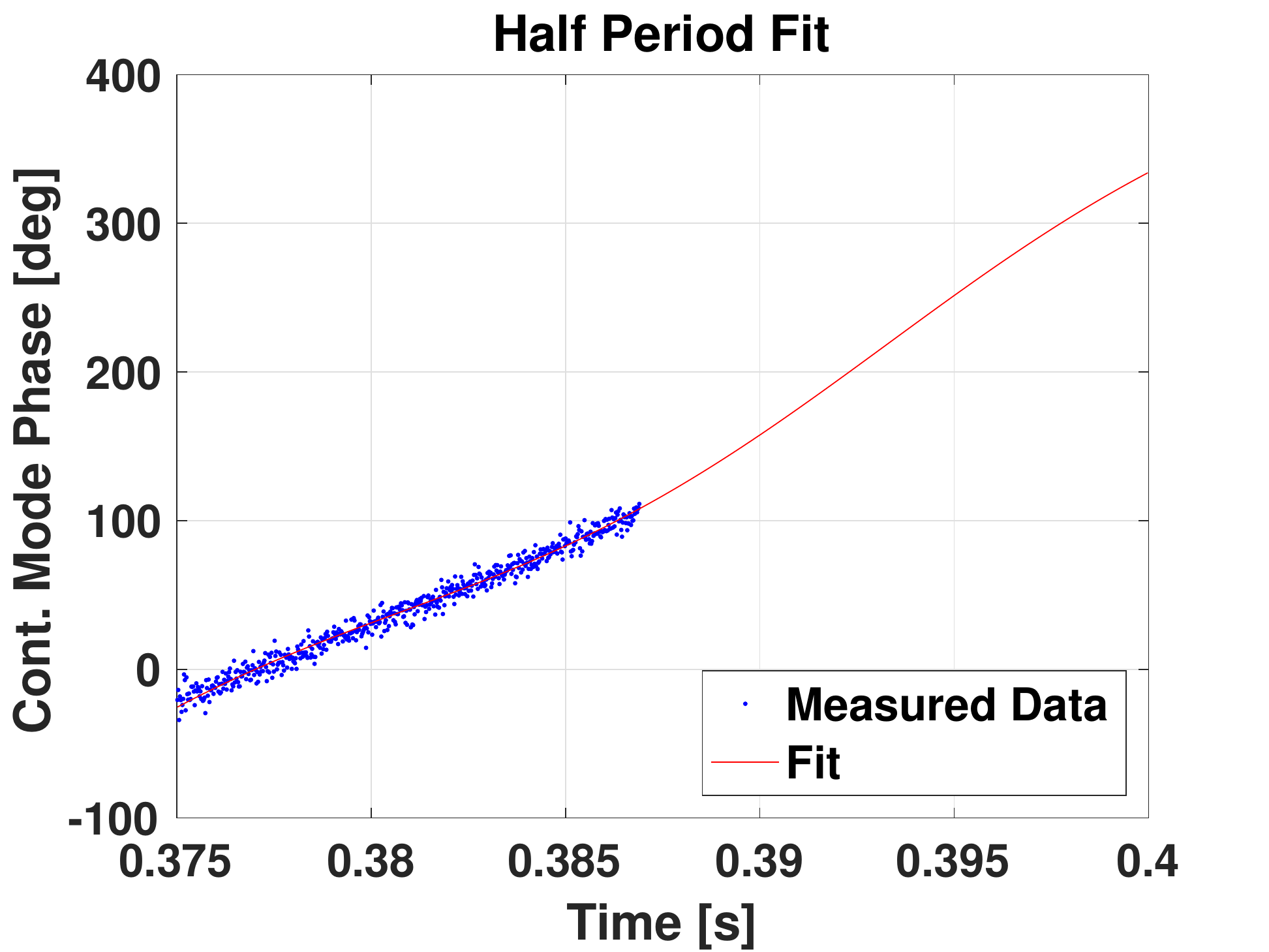}
	\caption{A half-period of data is sufficient to get a good estimate of the EF.}
	\label{fig:HalfPeriodFit}
\end{figure}

For faster convergence, the time per iteration can be shortened. 
It is necessary to accurately fit the non-uniform component of the fitting function, as it gives the error field amplitude and phase. 
Figure \ref{fig:HalfPeriodFit} shows that half a period of data is sufficient for a good fit using equation \ref{eqn:fittingFunction}.
This is supported by a symmetry argument, where oscillations of the mode rotation is mirrored at the EF phase.
Further reducing the measurement window to less than half a period, however, 
no long guarantees that at least one minimum or maximum of the sinusoidal oscillation will be found in the interval examined. 

\begin{figure}
	\includegraphics[width = 0.8 \columnwidth]{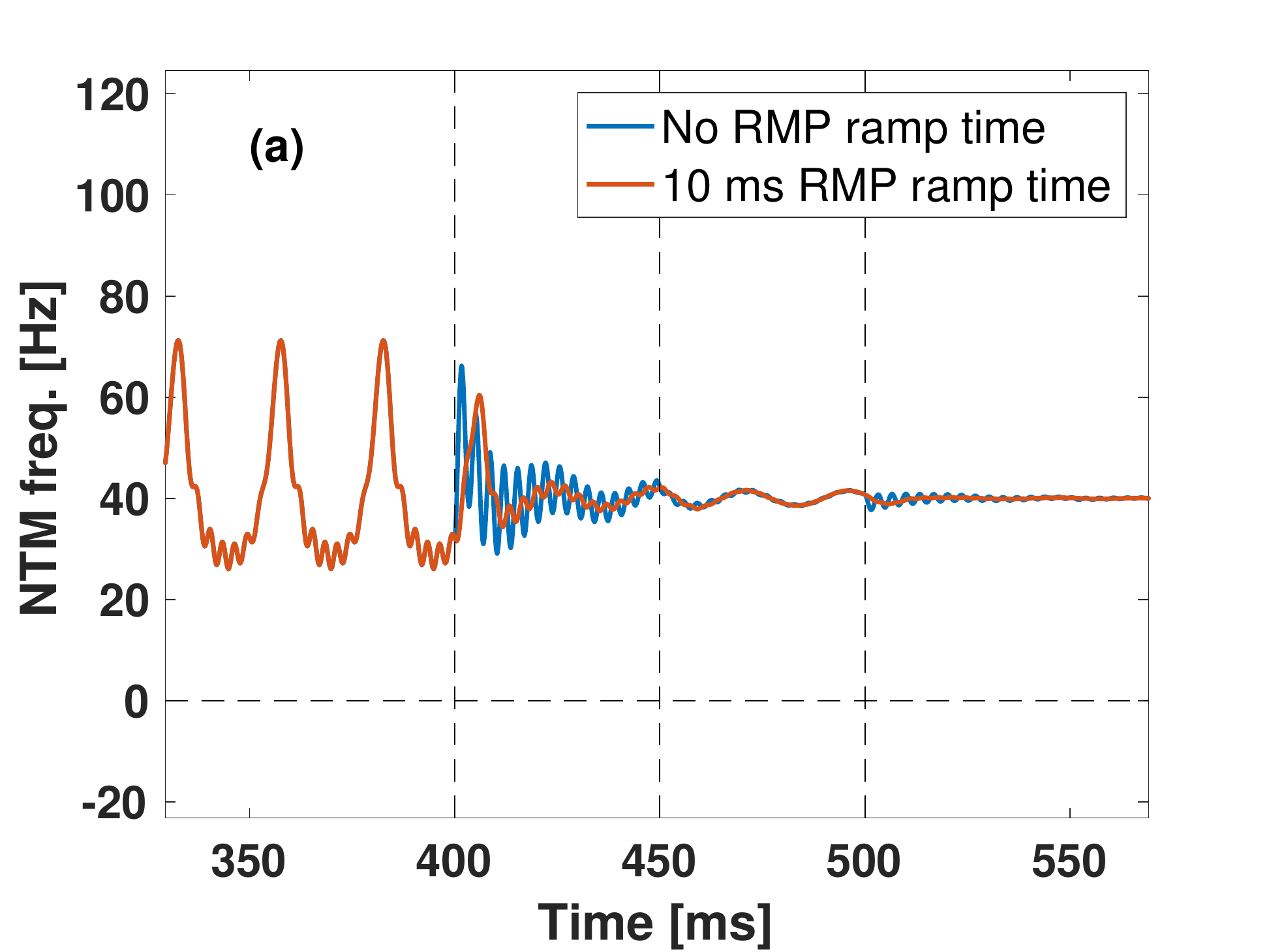}
	\includegraphics[width = 0.8 \columnwidth]{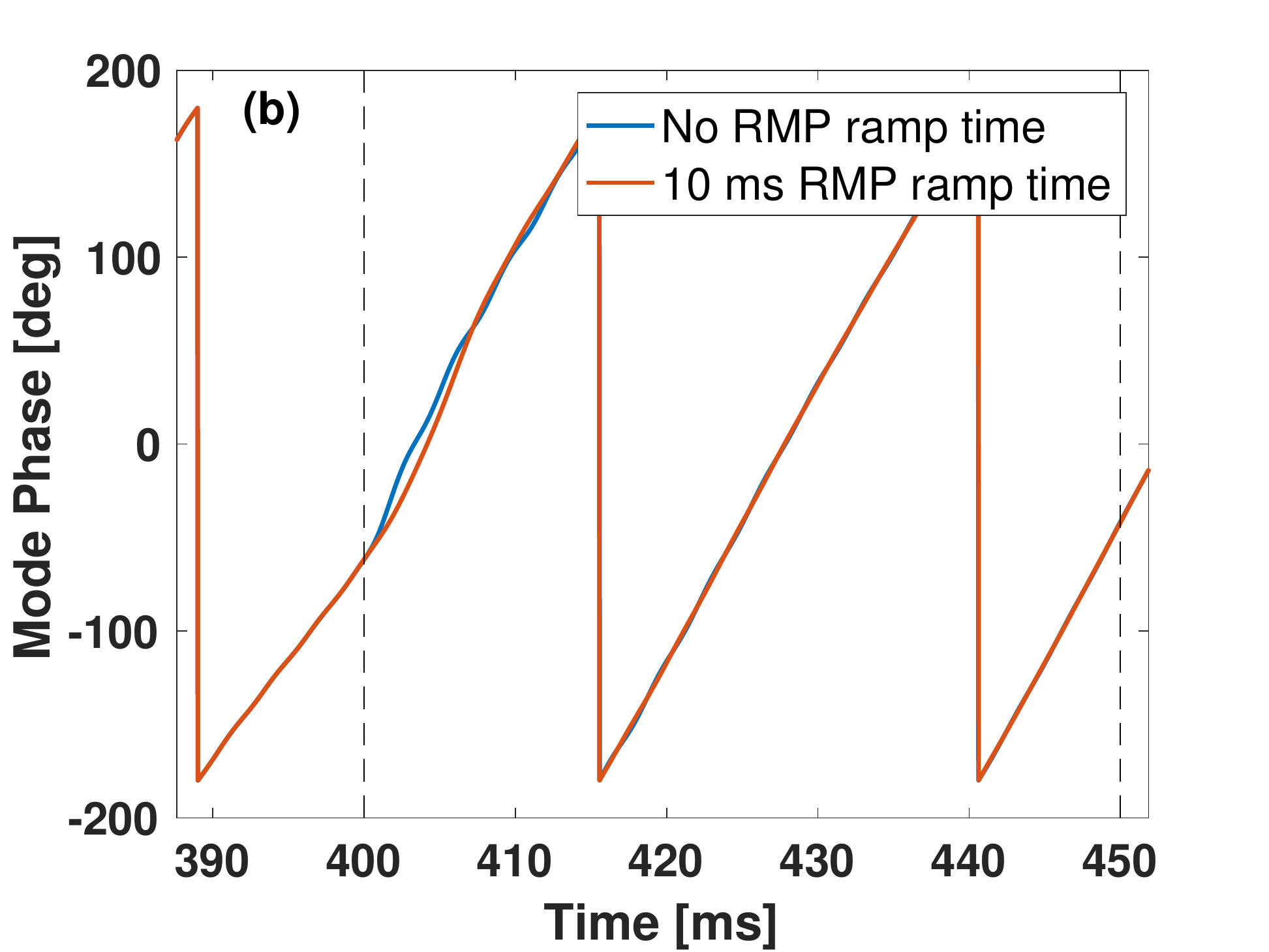}
	\caption{Ramping the correction over 10~ms prevents oscillations in island motion.}
	\label{fig:RMPramp}
\end{figure}

The settling period of the island motion can also be reduced by gradually applying the needed correction in the coils.
A step-function change (much faster than the wall time) in the applied currents induces strong eddy currents and causes unwanted oscillations in the rotation, as seen in figure \ref{fig:RMPramp}a.
While a slow ramp rate results in smoother motion, the trade-off is the longer time needed to reach the new EFC, which delays the measurement.
The improved mode phase behavior, shown in figure \ref{fig:RMPramp}b, reduces the settling time of the motion and allows the measurement window to begin earlier in the iteration period, 

While the two methods discussed above can shorten iteration periods, overly compressing each iteration reduces its performance and thus requires a larger number of iterations to achieve convergence. 
The fastest error field corrections needs an optimal balance between shortest time per iteration and smallest number of iterations.

\section{Conclusions}
\label{sec:conclusions}
Error fields have long been known to prevent high plasma performance.
Traditional methods of error field correction (EFC) typically involve a 4-discharge compass scan for a given configuration, 
then scaling the required correction for different plasma configurations.

The entrainment of magnetic islands by applied rotating magnetic fields 
has also long been known to have benefits for stability. 

Here, an iterative learning control (ILC) algorithm has been developed to correct EFs and improve the quality of mode entrainment in real time. 
This is achieved by optimizing the EFC as to maximize the uniformity or ``smoothness'' of island rotation.
In simulations, this controller has been shown to perform well 
regardless of the applied perturbation being larger or smaller than the 
initial EF. The only requisite is that the mode toroidal phase is 
measured in real time with a precision of about $\pm10^\circ$, which is fairly standard.


At the same time, by iteratively making the EF smaller and smaller, 
this method allows to ramp the island rotation-frequency to values 
that would not otherwise be accessible, paving the way to rotational 
stabilization, reliable disruption-avoidance and improved confinement.

\section*{Acknowledgements}
The authors would like to thank Ted Strait for the fruitful discussions and 
Nikolas Logan for his stimulating questions. 
This work was realized under DOE Grants DE-SC0008520, DE-SC0016372, and DE-FC02-04ER54698.

DISCLAIMER: This report was prepared as an account of work sponsored by an agency of the United States Government.  Neither the United States Government nor any agency thereof, nor any of their employees, makes any warranty, express or implied, or assumes any legal liability or responsibility for the accuracy, completeness, or usefulness of any information, apparatus, product, or process disclosed, or represents that its use would not infringe privately owned rights. Reference herein to any specific commercial product, process, or service by trade name, trademark, manufacturer, or otherwise, does not necessarily constitute or imply its endorsement, recommendation, or favoring by the United States Government or any agency thereof. The views and opinions of authors expressed herein do not necessarily state or reflect those of the United States Government or any agency thereof.


\printbibliography

\clearpage
\section*{Appendix: critical frequencies for loss of entrainment}
\label{sec:appendix}
Perfect entrainment entails uniform mode rotation with 
an inevitable but easy-to-calculate phase-lag 
with respect to the rotating RMP, due to the wall (section \ref{subsec:ConsidFast}). 
The rotation-frequency is constant even on short, sub-period timescales. 

Three different definitions of imperfect entrainment are discussed below, for completeness. 
Only the first, most stringent criterion was adopted in the body of the paper, because preventing that imperfection automatically prevents the other two. 

\subsection*{Diverging oscillations in rotation frequency}
As discussed, EFs introduce non-uniformities in the 
mode rotation. As a result, the rotation-frequency $f$ is not constant. Rather, it varies between two extremes within every rotation-period. Said otherwise, no 
torque-balance can be established,  in the 
presence of EFs, except for 2$n$ instants in every rotation-period. 
This is due to the EF torque oscillating on a sub-rotation timescale. 
As a consequence, the mode 
repetitively accelerates and decelerates, as if trying to align with the EF. 

\begin{figure}
	\includegraphics[width =  \columnwidth]{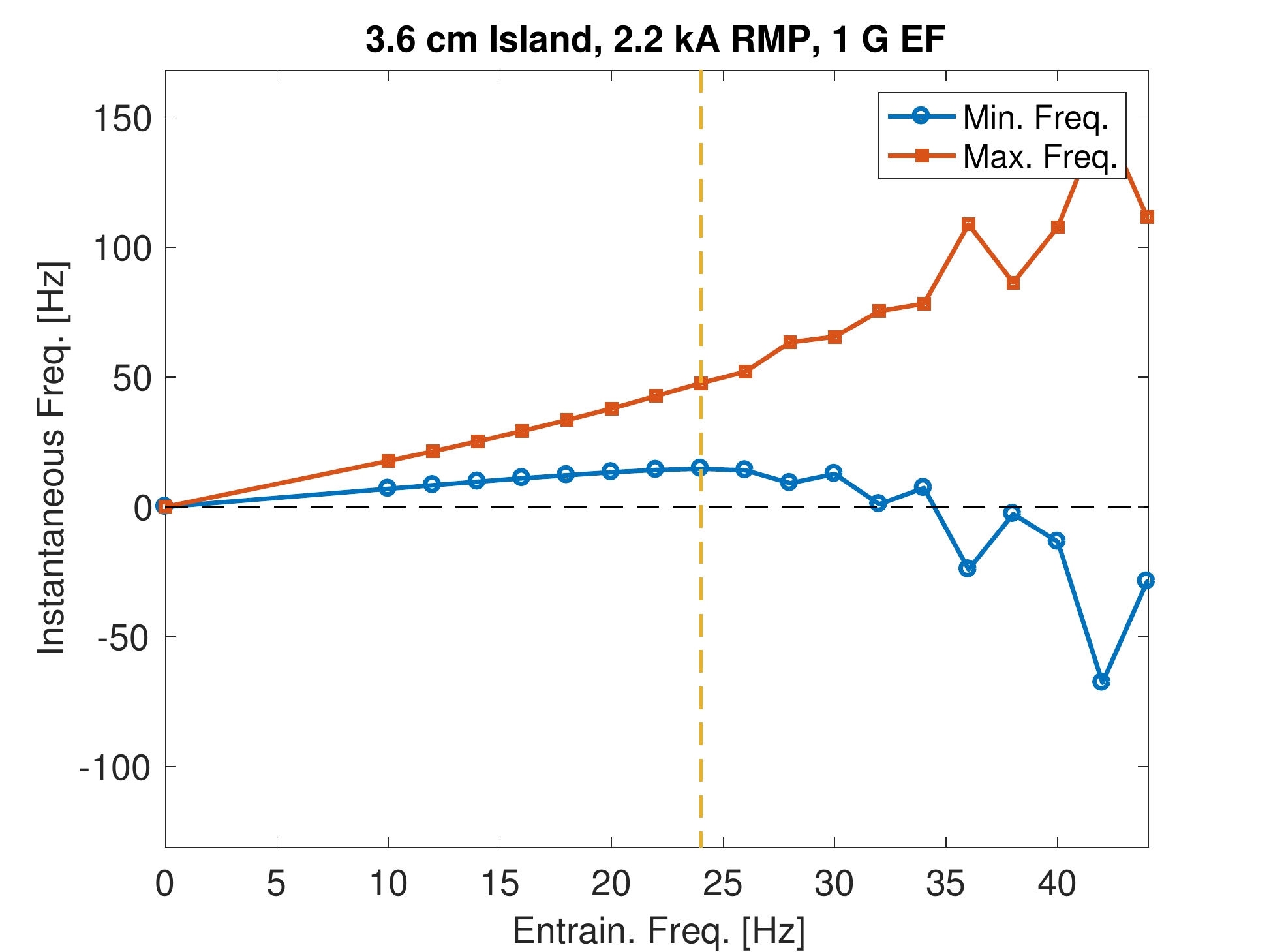}
	\caption{Minimum and maximum of instantaneous frequency of the mode during entrainment in the presence of EF.}
	\label{fig:divergeFreq}
\end{figure}

As the applied rotation-frequency $f_{RMP}$ increases, the instantaneous mode-frequency $f$ oscillates by larger and 
larger amounts $\pm \Delta f$. 
Beyond a certain value of $f_{RMP}$, $\Delta f$ is observed to diverge (figure \ref{fig:divergeFreq}). 
For brevity we call that value ``critical entrainment frequency''. 
However, it should be clarified that the mode can be entrained\textemdash in the sense that its rotation is magnetically sustained\textemdash 
even for higher $f_{RMP}$, and thus at higher $f\approx f_{RMP}$, albeit subject to large oscillations 
$\pm \Delta f$.  

Critical frequencies were plotted in figure \ref{fig:critFreqContours} as functions 
of the EF and RMP amplitudes, as well as of the island width. 
They are the results of time-resolved simulations of interactions between a rigid island, a resistive wall, a known error field, and a rotating RMP applied by 3D coils.

\subsection*{Sub-periods of counter-rotation}
\begin{figure}
	\includegraphics[width =  \columnwidth]{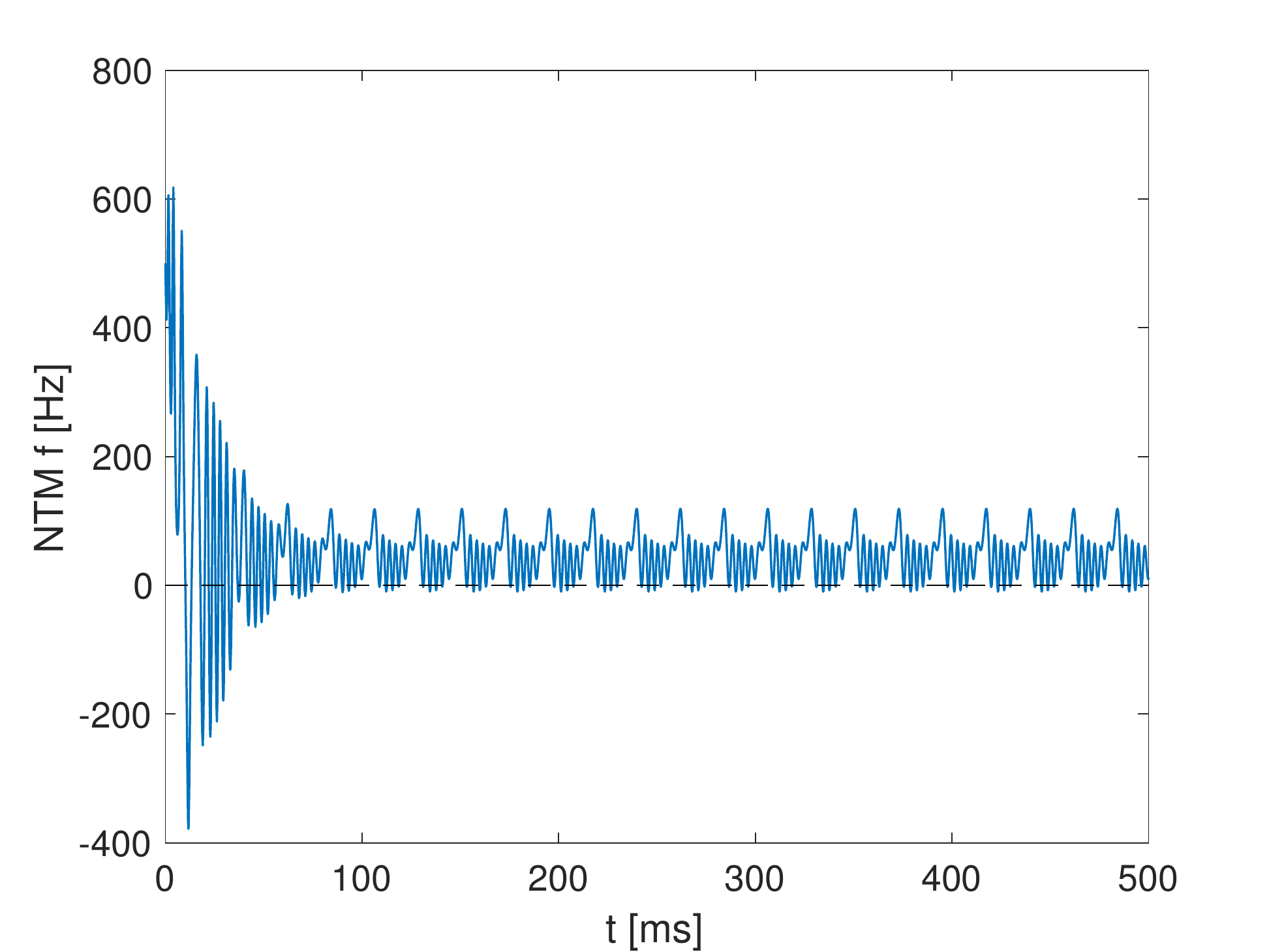}
		\includegraphics[width =  \columnwidth]{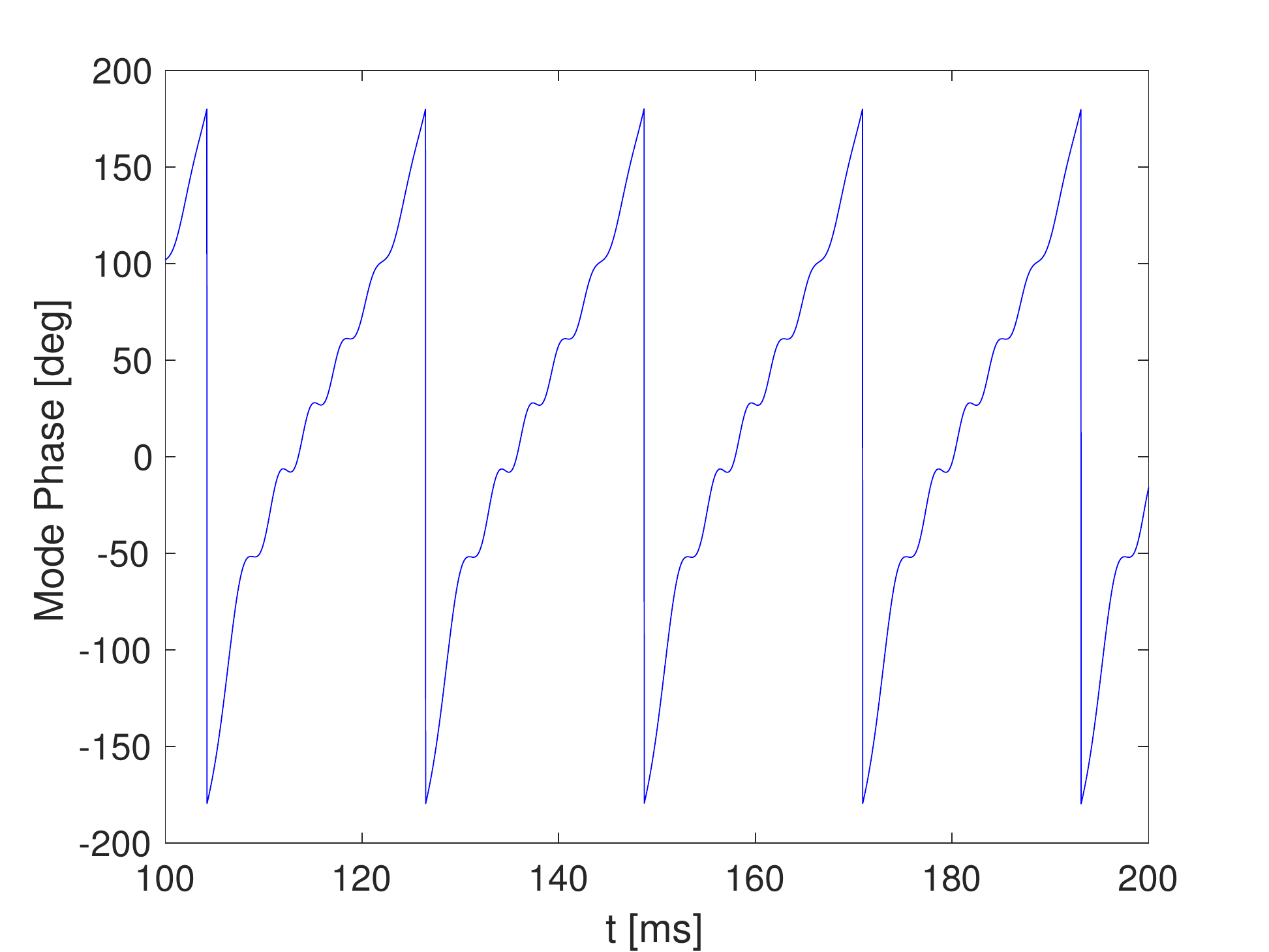}
	\caption{Example of mode rotation with temporary ``negative'' frequency.}
	\label{fig:reverseRot}
\end{figure}

As a consequence of the oscillations just discussed, $f$ can change sign for brief periods of time, 
shorter than a rotation period. 
That is, the mode can momentarily rotate opposite to the ``average'' direction of rotation 
(figure \ref{fig:reverseRot}). 

\begin{figure*}[t]
	\includegraphics[width = 0.33 \textwidth]{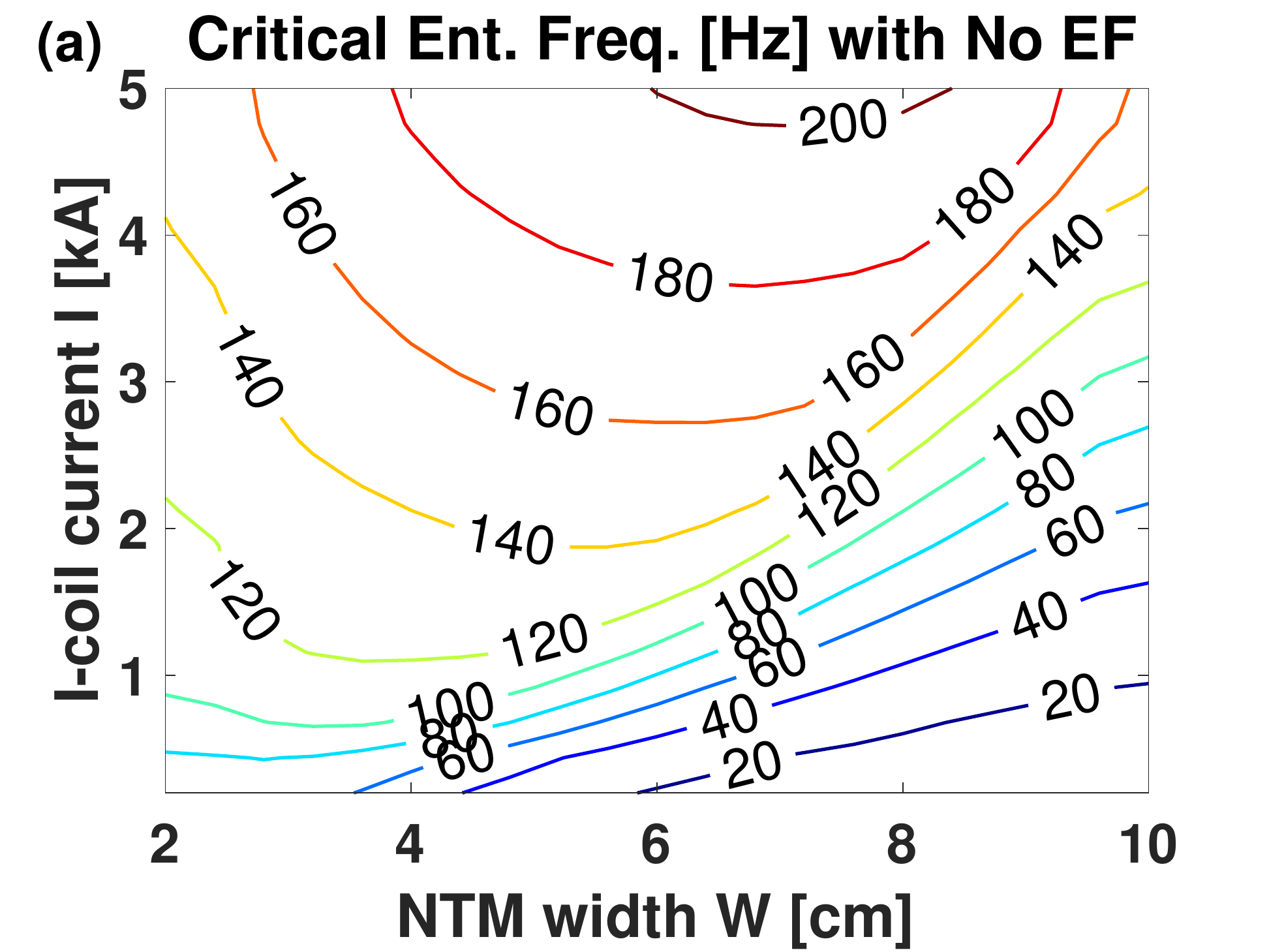}
	\includegraphics[width = 0.33 \textwidth]{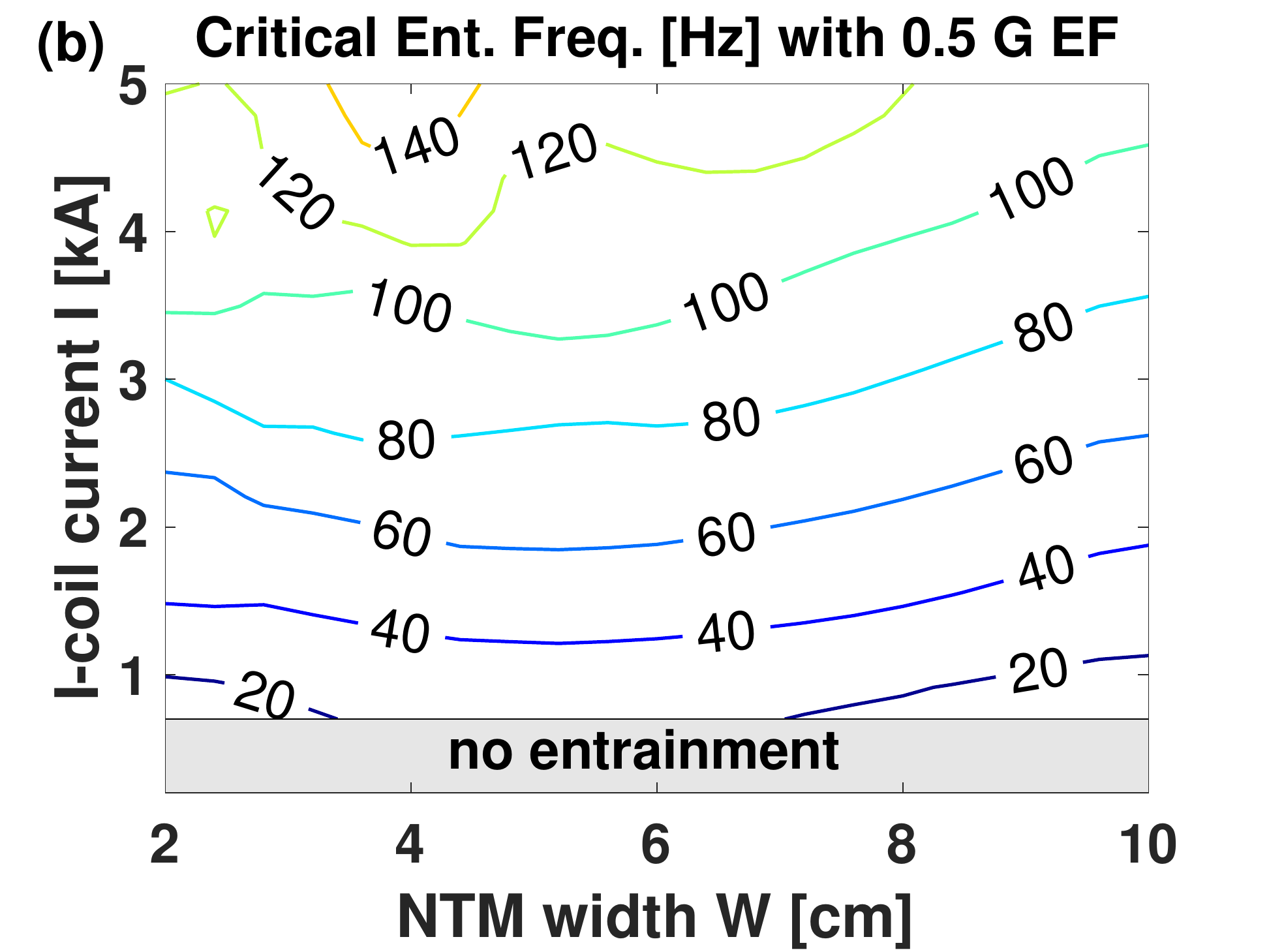}
	\includegraphics[width = 0.33 \textwidth]{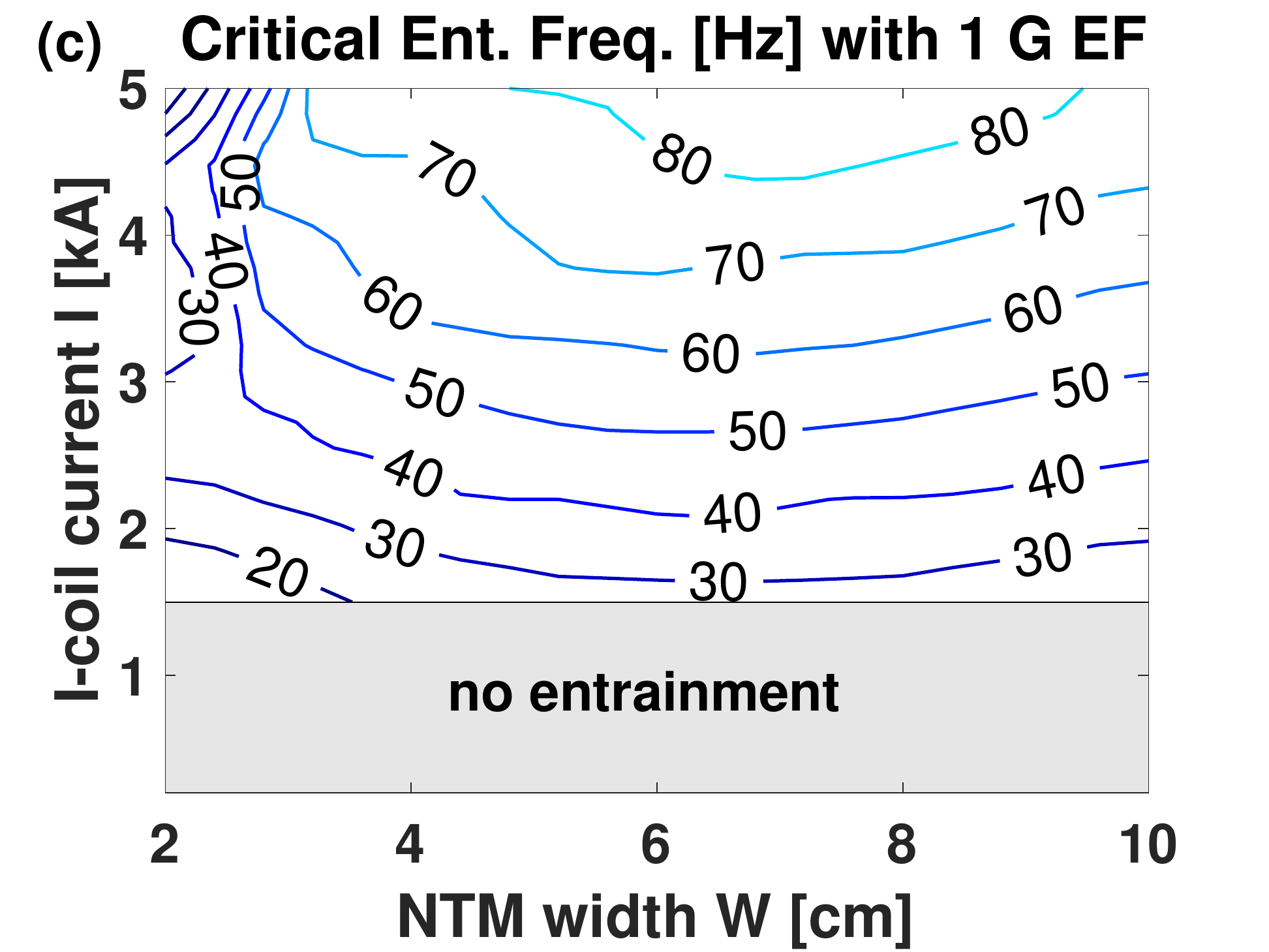}
	\caption{Critical entrainment frequency limits with and without EF, defined as having sub-periods of counter-rotation.}
	\label{fig:critFreqContoursMinFreqZero}
\end{figure*}

Figure \ref{fig:critFreqContoursMinFreqZero} shows the critical value of $f_{RMP}$, beyond which $f$ changes sign within a period, 
for similar RMP, island width, and EF strengths as in figure \ref{fig:critFreqContours}. 

\subsection*{Complete loss of entrainment}
In the two criteria considered so far, there was no {\em instantaneous} torque-balance sustained during one period. 
Despite that, the torque could be balanced in an {\em average}  sense, over a rotation period. 
Also, rotations were complete, in the sense of spanning all values of $\phi_{LM}$. 
This is the minimum requirement for most applications of entrainment, e.g. to diagnostics or to modulated 
ECCD stabilization.

At high enough $f_{RMP}$, that is not possible anymore: $\phi_{LM}$ oscillates in a  
range $<360^\circ$ (the turning points introduced in the last section are now global, not local)  
and we say that entrainment is completely lost. 

\begin{figure*}[t]
	\includegraphics[width = 0.33 \textwidth]{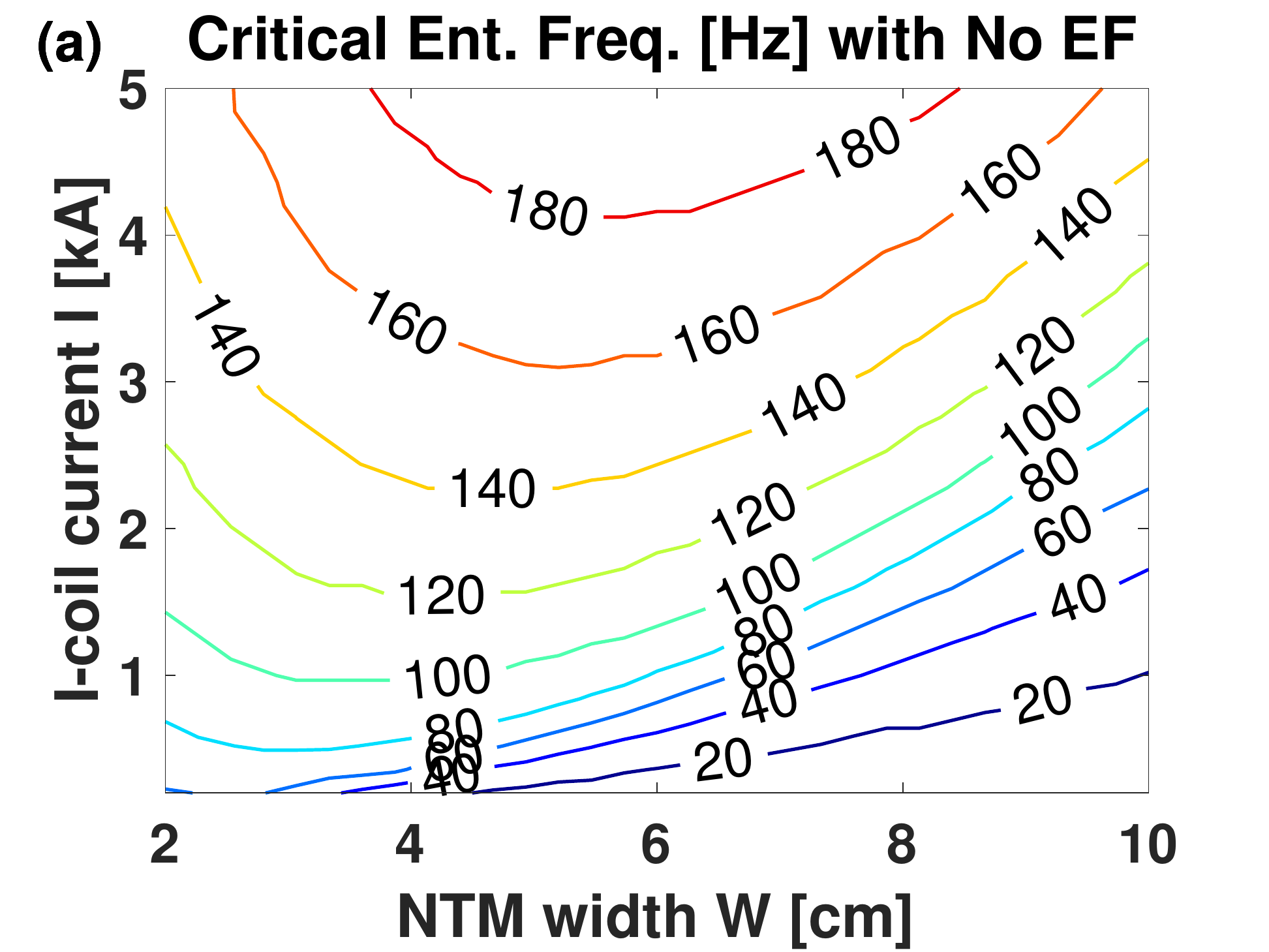}
	\includegraphics[width = 0.33 \textwidth]{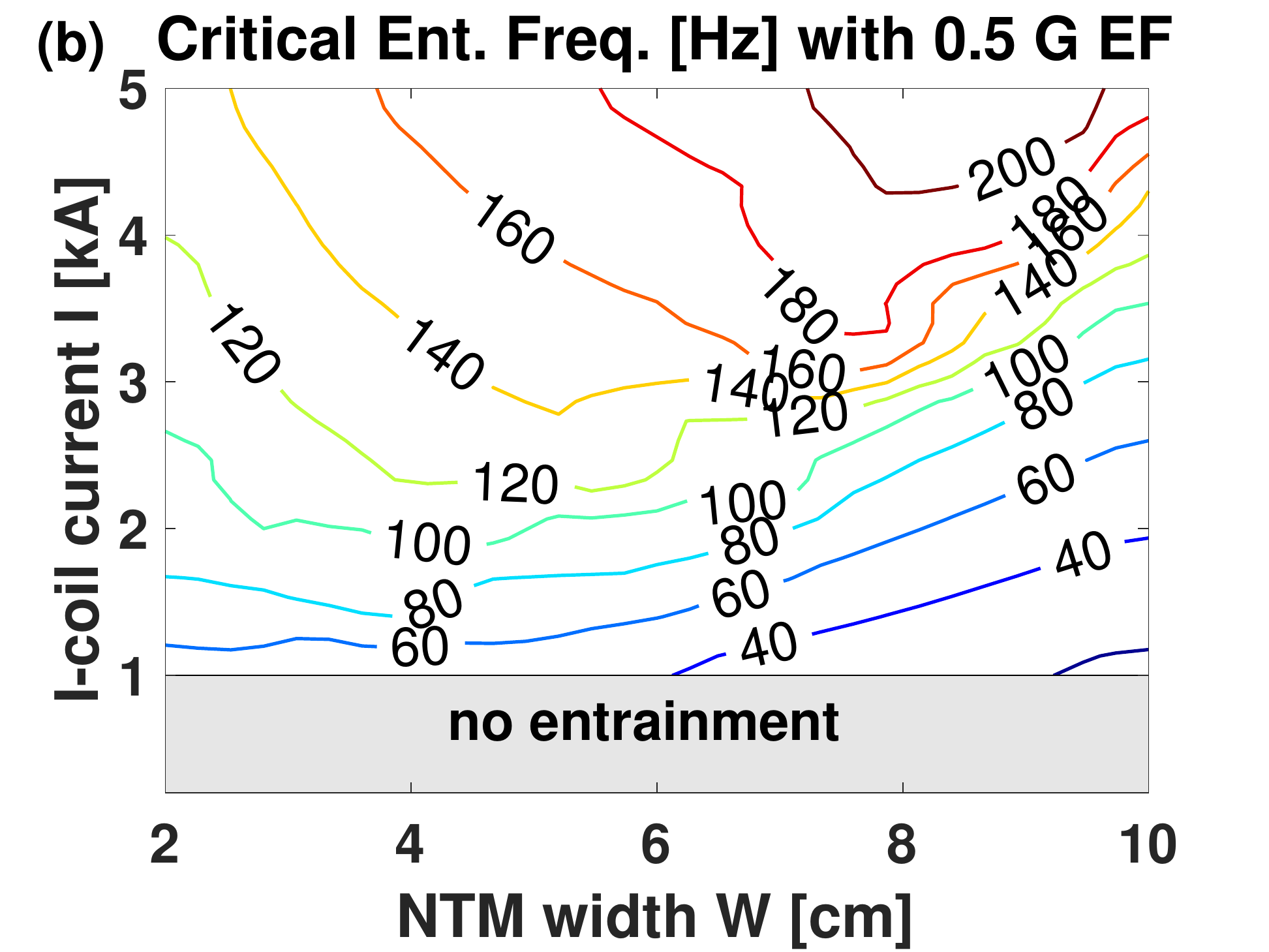}
	\includegraphics[width = 0.33 \textwidth]{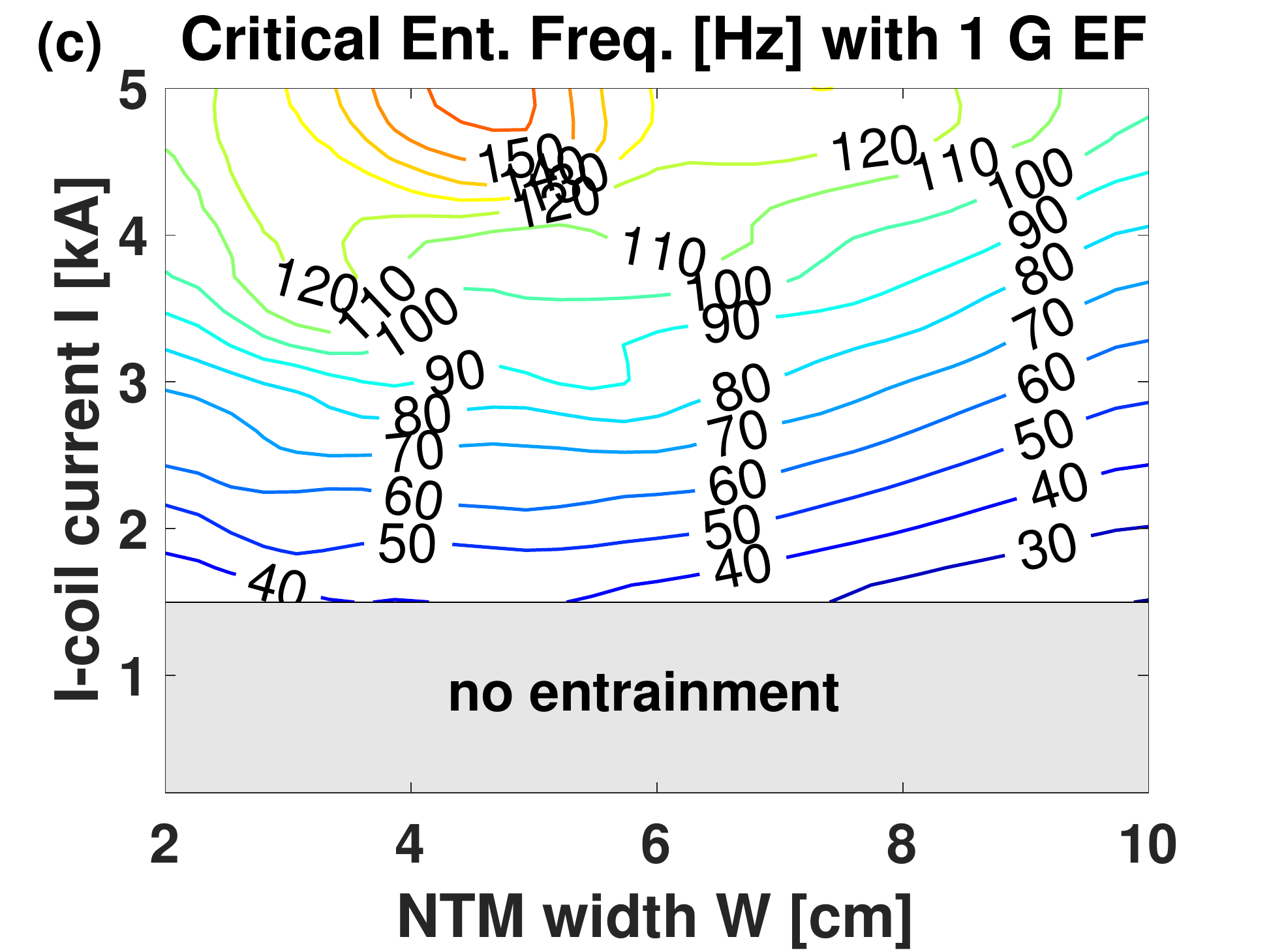}
	\caption{Critical entrainment frequency limits with and without EF, defined as having non-periodic, erractic motion.}
	\label{fig:critFreqContoursCompLoss}
\end{figure*}

Contours for this ``hard limit'' on entrainment were computed by means of time-resolved simulations of mode dynamics, similar to figure \ref{fig:critFreqContours} and \ref{fig:critFreqContoursMinFreqZero}, and are shown in figure \ref{fig:critFreqContoursCompLoss}. 

Due to the complex mutual dependence between island-dynamics and wall-currents, 
some simulations exhibited relatively long time-intervals of incomplete rotations (longer than a rotation period) alternated to several periods of full rotation. 
Those cases were categorized as complete losses of entrainment for the sake of preparing figure \ref{fig:critFreqContoursCompLoss}.

\end{document}